\documentclass{webofc}
\usepackage[varg]{txfonts}   
\usepackage{graphicx}  
\usepackage{dcolumn}   
\usepackage{bm}        
\usepackage{amssymb}   
\usepackage[caption=false]{subfig}
\usepackage{siunitx}

\DeclareSIUnit \s {\second}
\DeclareSIUnit \MB {\mega\byte}
\DeclareSIUnit \GB {\giga\byte}
\DeclareSIUnit \TB {\tera\byte}
\DeclareSIUnit \PB {\peta\byte}
\DeclareSIUnit \Mbps {\mega\bit/\s}
\DeclareSIUnit \Gbps {\giga\bit/\s}
\DeclareSIUnit \Tbps {\tera\bit/\s}
\DeclareSIUnit \Pbps {\peta\bit/\s}
\DeclareSIUnit \kton {\kilo\tonne} 
\DeclareSIUnit \kt {\kilo\tonne}
\DeclareSIUnit \Mt {\mega\tonne}
\DeclareSIUnit \eV {\electronvolt}
\DeclareSIUnit \keV {\kilo\electronvolt}
\DeclareSIUnit \MeV {\mega\electronvolt}
\DeclareSIUnit \GeV {\giga\electronvolt}
\DeclareSIUnit \m {\meter}
\DeclareSIUnit \cm {\centi\meter}
\DeclareSIUnit \in {\inchcommand}
\DeclareSIUnit \km {\kilo\meter}
\DeclareSIUnit \kV {\kilo\volt}
\DeclareSIUnit \kW {\kilo\watt}
\DeclareSIUnit \MW {\mega\watt}
\DeclareSIUnit \MHz {\mega\hertz}
\DeclareSIUnit \mrad {\milli\radian}
\DeclareSIUnit \year {year}
\DeclareSIUnit \POT {POT}
\DeclareSIUnit \sig {$\sigma$}
\DeclareSIUnit\parsec{pc}
\DeclareSIUnit\lightyear{ly}
\DeclareSIUnit\foot{ft}
\DeclareSIUnit\ft{ft}
\DeclareSIUnit \ppb{ppb}
\DeclareSIUnit \ppt{ppt}
\DeclareSIUnit \samples{S}

\newcommand{\refsec}[2]{Volume~\csname volnumber#1\endcsname \xspace Section~#2}
\newcommand{\refch}[2]{Volume~\csname volnumber#1\endcsname \xspace Chapter~#2}
\newcommand{\refinch}[2]{#2 in Volume~\csname volnumber#1\endcsname \xspace}

\newcommand{\numu}{\ensuremath{\nu_\mu}\xspace}
\newcommand{\nue}{\ensuremath{\nu_e}\xspace}

\def\argon40{${}^{40}$Ar}       
\def\Ar39{$^{39}$Ar}
\def\Cl40{$^{40}$Cl}
\def\K40{$^{40}$K}
\def\B8{$^{8}$B}

\def\fdfiducialmass{\SI{40}{\kt}\xspace}

\def\nominalmodsize{\SI{10}{kt}\xspace} 
\newcommand{\efield}{E field\xspace}

\newcommand{\threed}{3D\xspace}
\newcommand{\twod}{2D\xspace}
\newcommand{\phel}{photoelectron\xspace} 
\newcommand{\frfour}{FR-4\xspace} 

\newcommand{\lsim}{{\;\raise0.3ex\hbox{$<$\kern-0.75em\raise-1.1ex\hbox{$\sim$}}\;}}
\newcommand{\gsim}{{\;\raise0.3ex\hbox{$>$\kern-0.75em\raise-1.1ex\hbox{$\sim$}}\;}}
\newcommand{\beq}{\begin{equation}}
\newcommand{\eeq}{\end{equation}}
\newcommand{\bea}{\begin{eqnarray}}
\newcommand{\eea}{\end{eqnarray}}

\mathchardef\minus="002D

\newcommand{\rrt}[1]{}

\newcommand{\microboone}{MicroBooNE\xspace} 
\newcommand{\minerva}{MINERvA\xspace} 
\newcommand{\nova}{NOvA\xspace} 

\newcommand{\lartpc}{LArTPC\xspace}

\newcommand{\fnal}{Fermilab\xspace} 
\newcommand{\surf}{SURF\xspace}

\newcommand{\dual}{DP\xspace}

\newcommand{\single}{SP\xspace}

\newcommand{\lar}{LAr\xspace}

\usepackage[toc]{glossaries}
\makeglossaries

\newcommand{\dshort}[1]{\glsentrytext{#1}}  

\newcommand{\dword}[1]{\gls{#1}}

\newcommand{\newduneword}[3]{
    \newglossaryentry{#1}{
        text={#2},
        long={#2},
        name={\glsentrylong{#1}},
        first={\glsentryname{#1}},
        firstplural={\glsentrylong{#1}\glspluralsuffix},
        description={#3},
        sort={#2}
    }
}

\newcommand{\newduneabbrev}[4]{
  \newglossaryentry{#1}{
    text={#2},
    long={#3},
    shortplural={{#2}\glspluralsuffix},
    longplural={{#3}\glspluralsuffix{}},
    name={\glsentrylong{#1}{} (\glsentrytext{#1}{})},
    first={#3 (#2)},
    firstplural={#3\glspluralsuffix{} (\glsentrytext{#1}\glspluralsuffix{})},
    description={#4},
    sort={#2}
  }
}

\newcommand{\newduneabbrevs}[5]{
  \newglossaryentry{#1}{
    text={#2},
    long={#3},
    plural={#4},
    shortplural={{#2}\glspluralsuffix},
    longplural={#4},
    name={\glsentrylong{#1}{} (\glsentrytext{#1}{})},
    first={#3 (#2)},
    firstplural={#4 (\glsentrytext{#1}\glspluralsuffix{})},
    description={#5},
    sort={#2}    
  }
}

\newduneword{dword}{DUNE Word}{A term in the DUNE lexicon}

\newduneword{nasa}{NASA}{U.S. National Aereonautics and Space Administration}

\newduneabbrev{nd}{ND}{near detector}{Refers to the detector(s) 
 installed close to the neutrino source at \fnal }

\newduneabbrev{fd}{FD}{far detector}{The \fdfiducialmass fiducial mass DUNE detector, composed of four \nominalmodsize modules,  
  to be installed at the far site at \surf in
  Lead, SD, USA}

\newduneabbrev{sp}{SP}{single-phase}{Distinguishes one of the DUNE far detector technologies by the fact that it operates using argon in its liquid phase only}

\newduneabbrev{dp}{DP}{dual-phase}{Distinguishes one of the DUNE far detector technologies by the fact that it operates using argon 
 in both gas and liquid phases}

\newduneabbrev{pds}{PD system}{photon detection system}{The detector 
  subsystem sensitive to light produced in the \lar }

\newduneabbrev{hvs}{HVS}{high voltage system}{The detector 
  subsystem that provides the \gls{tpc} drift field}

\newduneabbrev{tpc}{TPC}{time projection chamber}{A type of particle detector that uses an \efield together with a sensitive volume of gas or liquid, e.g., \gls{lar}, to perform a \threed reconstruction of a particle trajectory or interaction. The activity is recorded by digitizing the waveforms of current
  induced on the anode as the distribution of ionization charge passes by
  or is collected on the electrode} 

\newduneabbrev{lartpc}{LArTPC}{liquid argon time-projection chamber}{A \gls{tpc} filled with liquid argon; 
the basis for the \gls{dune} \gls{fd} modules} 

\newduneabbrevs{apa}{APA}{anode plane assembly}{anode plane assemblies}{A unit of the \single
  detector module containing the elements sensitive to ionization in the \lar. 
  It contains two faces each of three planes of wires, and interfaces to the cold
  electronics and photon detection system} 

\newduneabbrev{awg}{AWG}{American wire gauge} {U.S. standard set of non-ferrous wire conductor sizes}

\newduneabbrev{ufer}{Ufer}{concrete encased electrode} {U.S. National Electrical Code grounding method refered to as Concrete Encased Electrode}

\newduneabbrev{cro}{CRO}{charge readout}{The system for detecting
  ionization charge distributions in a \dual detector module}

\newduneabbrev{lro}{LRO}{light readout}{The system for detecting
  scintillation photons in a \dual  detector module}

\newduneabbrev{shv}{SHV}{safe high voltage}{Type of bayonet mount
connector used on coaxial cables that has additional insulation 
compared to standard BNC and MHV connectors that makes it safer
for handling \gls{hv} by preventing accidental contact with the
live wire connector in an unmated connector or plug}

\newduneabbrev{fe}{FE}{front-end}{The front-end refers a point that is
  ``upstream'' of the data flow for a particular subsystem. 
  For example the \gls{sp} front-end electronics is where the cold electronics
  meet the sense wires of the TPC and the front-end \gls{daq} is where the
  \gls{daq} meets the output of the electronics}

\newduneabbrev{daqrou}{DAQ RU}{DAQ readout unit}{The first element in the data flow of the \gls{daq}}

\newduneabbrev{cots}{COTS}{commercial off-the-shelf}{Items, typically hardware such as 
computers, that may be purchased whole, without any custom design or fabrication and 
thus at normal consumer prices and availability}

\newduneabbrev{i2c}{I2C}{Inter-Integrated Circuit}{I$^2$C or I2C is a synchronous, 
multi-master, multi-slave, packet switched, single-ended, serial computer bus widely used 
for attaching lower-speed peripheral ICs to processors and microcontrollers in short-distance, 
intra-board communication} 

\newduneabbrev{spi}{SPI}{Serial Peripheral Interface}{The Serial Peripheral Interface is a 
synchronous serial communication interface specification used for short distance 
communication, primarily in embedded systems}

\newduneabbrev{miso}{MISO}{master in slave out}{The Master In Slave Out is a logic
signal on the \gls{spi} bus on which the data from the slave are transmitted once
a request from the master is received} 

\newduneabbrev{mosi}{MOSI}{master out slave in}{The Master Out Slave In is a logic
signal on the \gls{spi} bus on which the data from the master is transmitted} 

\newduneabbrev{uart}{UART}{Universal Asynchrous Receiver/Transmitter}{A universal 
asynchronous receiver-transmitter is a computer hardware device for asynchronous 
serial communication in which the data format and transmission speeds are configurable}

\newduneword{cr}{CR}{Capacitance-Resistance} 

\newduneword{dc}{DC}{direct coupling} 

\newduneword{ac}{AC}{capacitive coupling}  

\newduneabbrev{pll}{PLL}{Phase-Locked Loop}{A control system that generates an
output signal whose phase is related to the phase of an input signal}  

\newduneword{fifo}{FIFO}{First-In-First-Out} 

\newduneword{tsmc}{TSMC}{Taiwan Semiconductor Manufacturing Company}

\newduneword{saci}{SACI}{\gls{slac} \gls{asic} Control Interface}

\newduneword{om3}{OM3}{Type of multi-mode fiber optic cable, typically capable of \SI{10}{Gbps} data transmission at lengths up to \SI{300}{m}}

\newduneword{om4}{OM4}{Type of multi-mode fiber optic cable, typically capable of \SI{10}{Gbps} data transmission at lengths up to \SI{550}{m}}

\newduneword{qfp}{QFP}{Quad Flat Package} 

\newduneabbrev{ams}{AMS}{analog and mixed signal}{Verilog-AMS is a derivative of the Verilog hardware description language that includes analog and mixed-signal extensions (AMS) in order to define the behavior of analog and mixed-signal systems}

\newduneabbrev{hepa}{HEPA}{High Efficiency Particulate Air}{The High Efficiency Particulate Air filters are a type of air filter that remove \num{99.97}\% of particles that have a size greater han or equal to \SI{0.3}{$\mu$m}}  

\newduneabbrev{uvm}{UVM}{universal verification methodology}{The Universal Verification Methodology is a standardized methodology for verifying integrated circuit designs}   

\newduneword{lhc}{LHC}{Large Hadron Collider}

\newduneabbrev{lsb}{LSB}{least significant bit}{The bit with the lowest numerical value in a binary number}

\newduneabbrev{ldo}{LDO}{low-dropout regulator}{A low-dropout or LDO regulator is a \gls{dc} linear voltage regulator that can regulate the output voltage even when the supply voltage is very close to the output voltage}

\newduneabbrev{lpc}{LPC}{LHC Physics Cetner}{A Physics Center for LHC Physicists located at \dword{fnal}}

\newduneabbrev{irishep}{IRIS-HEP}{Institute for Research and Innovation in Software for High Energy Physics}{An NSF funded software institute. It aims to develop the state-of-the-art software cyberinfrastructure required for the challenges of data intensive scientific research at the High Luminosity Large Hadron Collider (HL-LHC) at CERN and other future experiments}

\newduneabbrev{adc}{ADC}{analog-to-digital converter}{A sampling of a voltage
  resulting in a discrete integer count corresponding in some way to
  the input}

\newduneabbrev{inl}{INL}{integral non-linearity}{A commonly used measure of performance in \glspl{adc}. It is the deviation between the ideal input threshold value and the measured threshold level of a certain output code}

\newduneabbrev{dnl}{DNL}{differential non-linearity}{A commonly used measure of performance in \glspl{adc}. The DNL error is defined as the difference between an actual step width and the ideal value of one \gls{lsb}}

\newduneword{pnp}{PNP}{Type of bipolar junction transistor consistning of a
layer of N-doped semiconductor sandwiched between two layers of P-doped material}

\newduneword{spice}{SPICE}{SPICE
(``Simulation Program with Integrated Circuit Emphasis'') is a general-purpose, 
open-source analog electronic circuit simulator. It is a program used in integrated 
circuit and board-level design to check the integrity of circuit designs and to 
predict circuit behavior}

\newduneabbrev{daq}{DAQ}{data acquisition}{The data acquisition system
  accepts data from the detector \gls{fe} electronics, buffers
  the data, performs a \gls{trigdecision}, builds events from the selected
  data and delivers the result to the offline \gls{diskbuffer}}

\newduneword{detmodule}{detector module}{The entire DUNE far detector is
  segmented into four modules, each with a nominal \SI{10}{\kton}
  fiducial mass}

\newduneword{detunit}{detector unit}{A 
portion of a \gls{detmodule} may be further partitioned into a number of similar parts.   For example the \gls{sp} \gls{tpc} 
is made up of \gls{apa}  units (and other elements)}

\newduneword{diskbuffer}{secondary DAQ buffer}{A secondary
  \dshort{daq} buffer holds a small subset of the full rate as
  selected by a \gls{trigcommand}. 
  This buffer also marks the interface with the DUNE Offline}

\newduneabbrev{om}{OM}{online monitoring}{Processes that run inside
  the \gls{daq} on data ``in flight,'' specifically before landing on the
  offline disk buffer, and that provide feedback on the operation of
  the \gls{daq} itself and the general health of the data it is marshalling}

\newduneabbrev{dqm}{DQM}{data quality monitoring}{Analysis of the raw
  data to monitor the integrity of the data and the performance of the
  detectors and their electronics. This type of monitoring may be
  performed in real time, within the \gls{daq} system, or in later
  stages of processing, using disk files as input}

\newduneword{dumpbuffer}{DAQ dump buffer}{This \gls{daq} buffer
  accepts a high-rate data stream, in aggregate, from an associated
  portion of a \gls{detmodule} sufficient to collect all data likely relevant to
  a potential \gls{snb}}

\newduneabbrev{etl}{ETL}{external trigger logic}{Trigger processing
  that consumes \gls{detmodule} level \gls{trignote} information
  and other global sources of trigger input and emits
  \gls{trigcommand} information back to the \glspl{mtl}}
\newduneabbrev{daqeti}{ETI}{external trigger interface}{Interface between \glspl{mtl} and external source and sinks of relevant trigger information}

\newduneword{trignote}{trigger notification}{Information provided by
  \gls{mtl} to \gls{etl} about \gls{trigdecision} 
  processing}

\newduneword{trigprimitive}{trigger primitive}{Information derived by
  the \gls{daq} \gls{fe} hardware that describes a region of space (e.g.,
  one or several neighboring channels) and time (e.g., a contiguous set
  of \gls{adc} sample ticks) associated with some activity}

\newduneword{externtrigger}{external trigger candidate}{Information
  provided to the \gls{mtl} about events external to a
  \gls{detmodule} so that it may be considered in forming
  \glspl{trigcommand}}

\newduneabbrev{daqoob}{OOB dispatcher}{out-of-band trigger command
  dispatcher}{This component is responsible for dispatching a \gls{snb} dump
  command to all \glspl{daqfer} in the \gls{detmodule}}

\newduneabbrev{mtl}{MTL}{module trigger logic}{Trigger processing
  that consumes \gls{detunit} level \gls{trigcommand} information
  and emits \glspl{trigcommand}. 
  It provides the \gls{etl} with \glspl{trignote} and receives back any
  \glspl{externtrigger}}

\newduneword{octant}{octant}{Any of the eight parts into which 4$\pi$
  is divided by three mutually perpendicular axes. 
  In particular in referencing the value for the mixing angle
  $\theta_{23}$}

\newduneword{trigcandidate}{trigger candidate}{Summary information derived
  from the full data stream and representing a contribution toward
  forming a \gls{trigdecision}}

\newduneword{trigcommand}{trigger command}{Information derived from
  one or more \glspl{trigcandidate}  that directs elements of the
  \gls{detmodule} to read out a portion of the data stream}

\newduneabbrev{tcm}{TCM}{trigger command message}{A message flowing
  down the trigger hierarchy from global to local context.  Also see \gls{tpm}}

\newduneabbrev{mlt}{MLT}{module level trigger}{The \gls{daq} component responsible for producing a \gls{trigdecision} that will be used to command the readout of a detector module}

\newduneword{trigdecision}{trigger decision}{The process by which
  \glspl{trigcandidate} are converted into \glspl{trigcommand}}

\newduneabbrev{tpm}{TPM}{trigger primitive message}{A message flowing
  up the trigger hierarchy from local to global context.  Also see \gls{tcm}}

\newduneabbrev{ipc}{IPC}{inter-process communication}{A system for software elements to exchange information between threads, local processes or across a data network.  An IPC system is typically specified in terms of protocols  composed of message types and their associated data schema}

\newduneword{daqdispre}{discovery and presence}{As used in the context of the \gls{ipc}, a system that provides mechanisms for a node on a communication network to learn of the existence of peers and their identity (discovery) as well as determine if they are currently operational or have become unresponsive (presence)}

\newduneabbrev{pubsub}{PUB/SUB}{publish-subscribe communication pattern}{An \gls{ipc} communication pattern where one element, the publisher, sends data to all connected elements, the subscribers.  Each subscriber may connect to multiple publishers.  A variant is PUB/SUB with topics where a subscriber may register an identifier, the topic, to limit the information received to just an associated subset}

\newduneabbrev{eb}{EB}{event builder}{A software agent that executes \glspl{trigcommand}  for one  \gls{detmodule} by reading out the requested data}

\newduneabbrev{daqdfo}{DFO}{data flow orchestrator}{The process by which trigger commands are executed in parallel and asynchronous manner by the back-end output subsystem of the \gls{daq}}

\newduneabbrev{daqubi}{UBI}{upstream DAQ buffer interface}{The process which provides read-only access to data residing in the upstream \gls{daq} buffers to processes on the network}

\newduneabbrev{cob}{COB}{cluster on board}{An ATCA motherboard housing four RCEs}

\newduneabbrev{rce}{RCE}{reconfigurable computing element}{Data processor located outside of the cryostat on a \gls{cob} that contains \gls{fpga}, RAM and \gls{ssd} resources, responsible for buffering data, producing trigger primitives, responding to triggered requests for data and synching \gls{snb} dumps}

\newduneabbrev{bow}{BOW}{Bump On Wire}{A working name for the front-end readout computing elements used in the nominal \gls{daq} design to interface the \dual  crates to the \gls{daq} front-end computers}

\newduneabbrev{atca}{ATCA}{Advanced Telecommunications Computing
  Architecture}{An advanced computer architecture specification developed for the telecommunications, military, and aerospace industries that incorporates the latest trends in  high-speed interconnect technologies, next-generation processors, and improved reliability, availability and serviceability} 

\newduneabbrev{utca}{$\mu$TCA}{Micro Telecommunications Computing Architecture}{The computer architecture specification followed by the crates that house charge and light readout electronics in the \gls{dpmod}} 

\newduneabbrev{udp}{UDP}{user datagram protocol}{A simple,
  connectionless Internet protocol that supports data integrity
  checksums, requires no handshaking, and does not guarantee packet delivery}

\newduneabbrev{amc}{AMC}{advanced mezzanine card}{Holds digitizing
  electronics and lives in \gls{utca} crates}

\newduneabbrev{rf}{RF}{radio frequency}{Electromagnetic emissions
  that are within the (radio) frequency band of sensitivity of the detector
  electronics}

\newduneabbrev{fpga}{FPGA}{field programmable gate array}{An
integrated circuit technology that allows the hardware to be reconfigured to
execute different algorithms after its manufacture and deployment}

\newduneabbrev{fmc}{FMC}{FPGA mezzanine card}{Boards holding \glspl{fpga} and other integrated circuitry that attach to a motherboard}

\newduneabbrev{felix}{FELIX}{Front-End Link eXchange}{A
  high-throughput interface between \gls{fe} and trigger electronics
  and the standard PCIe computer bus}

\newduneword{daqpart}{DAQ partition}{A cohesive and
 coherent collection of \gls{daq} hardware and software working together to trigger and read out some portion of one detector module; it consists of an integral number of \glspl{daqfrag}. 
 Multiple \gls{daq} partitions may operate simultaneously, but each instance operates independently}

\newduneabbrev{fec}{DAQ FEC}{DAQ front-end computer}{The portion of one
  \gls{daqpart} that hosts the \gls{daqdr}, \gls{daqbuf} and
  \gls{daqds}.  It hosts the \gls{daqfer} and corresponding portion of the \gls{daqbuf}}

\newduneword{daqfrag}{DAQ front-end fragment}{The portion of one
  \gls{daqpart} relating to a single \gls{fec} and corresponding to an
  integral number of \glspl{detunit}.  See also \gls{datafrag}}

\newduneword{datafrag}{data fragment}{A block of data read out from a single \gls{daqfrag} that
span a contiguous period of time as requested by a \gls{trigcommand}}

\newduneabbrev{daqfer}{FER}{DAQ front-end readout}{The portion of a
  \gls{daqfrag} that accepts data from the detector electronics and
  provides it to the \gls{fec}}

\newduneabbrev{daqdr}{DDR}{DAQ data receiver}{The portion of the
  \gls{daqfrag} that accepts data from the \gls{daqfer}, emits
  trigger candidates produced from the input trigger primitives, and
  forwards the full data stream to the \gls{daqbuf}}

\newduneword{daqbuf}{DAQ primary buffer}{The portion
  of the \gls{daqfrag} that accepts full data stream from the
  corresponding \gls{detunit} and retains it sufficiently long for it
  to be available to produce a \gls{datafrag}}

\newduneword{daqds}{data selector}{The portion of the \gls{daqfrag}
  that accepts \glspl{trigcommand} and returns the corresponding
  \gls{datafrag}.  Not to be confused with \gls{daqdsn}}

\newduneword{daqdsn}{data selection}{The process of forming a trigger decision for selecting a subset of detector data for output by the \gls{daq} from the content of the detector data itself.  Not to be confused with \gls{daqds}}

\newduneabbrev{daqros}{DAQ RO}{DAQ readout subsystem}{The subsystem of the \gls{daq} for accepting and buffering data input from detector electronics}

\newduneabbrev{daqdss}{DAQ DS}{DAQ data selection subsystem}{The subsystem of the \gls{daq} responsible for forming a trigger decision based on a portion of the input data stream.  The majority subset of the \gls{daqtrs}}

\newduneabbrev{daqtrs}{DAQ TS}{DAQ trigger subsystem}{The subsystem of the \gls{daq} responsible for forming a trigger decision}

\newduneabbrev{daqbes}{DAQ BE}{DAQ back-end subsystem}{The portion of the \gls{daq} that is generally toward its output end.  It is responsible for accepting and executing trigger commands and marshaling the data they address to output storage buffers}

\newduneabbrev{daqtss}{DAQ TSS}{DAQ timing and synchronization subsystem}{The portion of the \gls{daq} that provides for timing and synchronization to various components}

\newduneabbrev{femb}{FEMB}{front-end mother board}{Refers a unit of
  the \gls{sp} \gls{ce} that contains the \gls{fe} amplifier
  and \gls{adc} \glspl{asic} covering 128 channels}

\newduneword{asic}{ASIC}{application-specific integrated circuit}

\newduneword{lv}{LV}{low voltage}

\newduneabbrev{iceberg}{ICEBERG}{ICEBERG R\&D cryostat and electronics}{Integrated Cryostat and Electronics Built for Experimental Research Goals:
a new double-walled cryostat built and installed at \gls{fnal} 
for liquid argon detector R\&D and for testing of DUNE detector components}

\newduneword{coldadc}{ColdADC}{A newly developed 16-channels \gls{asic} providing analog to digital conversion}

\newduneword{coldata}{COLDATA}{A 64-channel control and communications \gls{asic}}

\newduneword{cryo}{CRYO}{Integrated ASIC including \gls{fe} circuitry providing signal amplification and pulse shaping, analog to digital conversion, and control and communication functionalities for 64 channels}

\newduneword{larasic}{LArASIC}{A 16-channel \gls{fe} \gls{asic} that provides signal amplification and pulse shaping}

\newduneword{cmos}{CMOS}{Complementary metal-oxide-semiconductor}

\newduneabbrev{enc}{ENC}{equivalent noise charge}{The equivalent noise charge is the input charge that corresponds to a 
$\gls{snr}=1$}

\newduneword{sar}{SAR}{successive approximation register}

\newduneword{protodune}{ProtoDUNE}{Either of the two DUNE prototype detectors constructed at \gls{cern}. 
  One prototype implements \gls{sp} technology and the other \gls{dp}}
  
\newduneword{protodune2}{ProtoDUNE-2}{The second run of a \gls{protodune} detector}

\newduneword{pdsp}{ProtoDUNE-SP}{The \gls{sp} \gls{protodune} detector at \gls{cern}}

\newduneword{pddp}{ProtoDUNE-DP}{The \gls{dp} \gls{protodune} detector at \gls{cern}}

\newduneword{wa105}{WA105 DP demonstrator}{The \SI[product-units=power]{3x1x1}{m} WA105 \gls{dp} prototype detector at \gls{cern}}

\newduneword{rawevent}{DAQ event block}{The unit of data output by the
  \gls{daq}.  
  It contains trigger and detector data spanning a unique, contiguous
  time period and a subset of the detector channels}

\newduneabbrev{ssd}{SSD}{solid-state disk}{Any storage device that
  may provide sufficient write throughput to receive, both collectively and
  distributed, the sustained full rate of data from a \gls{detmodule}
  for many seconds}
\newduneabbrev{nvme}{NVMe}{Non-volatile memory express}{A specification for an interface to storage media attached via PCIe}

\newduneabbrev{hlt}{HLT}{high-level trigger}{This is actually a filter applied to data that has been triggered and aggregated in order to further reduce or characterize it}

\newduneabbrev{pid}{PID}{particle ID}{Particle identification}

\newduneword{readout window}{readout window}{A fixed, atomic and
  continuous period of time over which data from a \gls{detmodule}, in
  whole or in part, is recorded. 
  This period may differ based on the trigger that initiated the
  readout}

\newduneabbrev{zs}{ZS}{zero-suppression}{Used to delete some portion of a
  data stream that does not significantly deviate from zero or
  intrinsic noise levels. 
  It may be applied at different granularity from per-channel to per
  \gls{detunit}}

\newduneabbrev{rc}{RC}{run control}{The system for configuring,
  starting and terminating the \gls{daq}}

\newduneword{r-c}{RC}{resistive-capacitive (circuit)}

\newduneabbrev{daqccm}{CCM}{DAQ control, configuration and monitoring subsystem}{A system for controlling, configuring and monitoring other systems in particular those that make up the \gls{daq} where the CCM encompasses \gls{rc}}

\newduneword{daqrun}{DAQ run}{A period of time over which relevant data taking conditions and \gls{daq} configuration are asserted to be unchanged. 
  Multiple \gls{daq} runs may occur simultaneously when multiple \glspl{daqpart} are active. 
  This term should not be confused with DUNE experiment or beam ``runs'' that typically span many \gls{daq} runs}
\newduneword{daqrunnum}{DAQ run number}{A monotonically increasing count that uniquely and globally identifies a \gls{daqrun}}

\newduneabbrev{snb}{SNB}{supernova neutrino burst}{A prompt 
  increase in the flux of low-energy neutrinos emitted in the first few seconds of a core-collapse supernova.  It can also refer to a trigger command type that may be due to this phenomenon,
  or detector conditions that mimic its interaction signature}

\newduneabbrev{snble}{SNB/LE}{supernova neutrino burst and low
  energy}{Supernova neutrino burst and low-energy physics program}

\newduneabbrev{snews}{SNEWS}{SuperNova Early Warning System}{A global
  supernova neutrino burst trigger formed by a coincidence of \gls{snb} 
  triggers collected from participating experiments}

\newduneabbrev{pps}{1PPS signal}{one-pulse-per-second signal}{An
  electrical signal with a fast rise time and that arrives in real
  time with a precise period of one second}

\newduneabbrev{sls}{SLS}{spill location system}{A system residing at
  the DUNE far detector site that provides information, possibly
  predictive, indicating periods of time when neutrinos are being
  produced by the \fnal Main Injector beam spills}

\newduneabbrev{wib}{WIB}{warm interface board}{Digital electronics
  situated just outside the \gls{sp} cryostat that receives digital data
  from the \glspl{femb} over cold copper connections and sends it to the \gls{rce}
  \gls{fe} readout hardware}

\newduneabbrev{gps}{GPS}{Global Positioning System}{A satellite-based system that provides a highly accurate \gls{pps} that may be used to synchronize clocks and determine location}

\newduneabbrev{ntp}{NTP}{Network Time Protocol}{A networking protocol that allows synchronizing of clocks to within a few \si{\milli\second} of a time standard on a local network and within a few tens of \si{\milli\second} over the Internet} 

\newduneabbrev{ptproto}{PTP}{Precision Time Protocol}{A networking protocol that allows synchronizing of clocks to within a few \si{\micro\second} of a time standard on a local network} 

\newduneabbrev{irig}{IRIG}{inter-range instrumentation group}{A standards body that defined a time-code standard for transferring timing information}

\newduneabbrev{nic}{NIC}{network interface controller}{Hardware for controlling the interface to a communication network.  Typically, one that obeys the Ethernet protocol}

\newduneabbrev{wiec}{WIEC}{warm interface electronics crate}{Crates mounted on the signal flanges that contain the \glspl{wib}}

\newduneabbrev{ptc}{PTC}{power and timing card}{Cards that provide further processing and distribution of the signals entering and exiting the \gls{sp} cryostat}

\newduneabbrev{ptb}{PTB}{power and timing backplane}{Backplane used to connect the \gls{wib}s and the \gls{ptc}s on the \gls{wiec}. Also connects the \gls{ce} flange on the cryostat penetration}

\newduneabbrev{sipm}{SiPM}{silicon photomultiplier}{A solid-state
  avalanche photodiode sensitive to single \phel signals}

\newduneabbrev{cisc}{CISC}{cryogenic instrumentation and slow controls}{Includes equipment to monitor all detector  components and  \gls{lar} quality and behavior, and provides a control system for many of the detector components}

\newduneword{fte}{FTE}{full-time equivalent. A unit of labor
  for the project. One year of work from one person}

\newduneword{art}{art}{A software framework implementing an
  event-based execution paradigm} 

\newduneabbrev{sam}{SAM}{sequential
  access via metadata}{A data-handling system to store and retrieve
  files and associated metadata, including a complete record of the
  processing that has used the files}

\newduneword{artdaq}{artdaq}{A data acquisition toolkit for data transfer, aggregation and processing}

\newduneword{beamline}{beamline}{A sequence of control and monitoring devices used for the formation of a directed collection of particles}
\newduneabbrev{cdr}{CDR}{conceptual design report}{A formal project
  document 
   that describes the experiment
  at a conceptual level}

\newduneabbrev{cf}{CF}{conventional facilities}{Pertaining to
  construction and operation of buildings and conventional infrastructure, and for \gls{lbnf-dune}, CF includes the excavation caverns}

\newduneabbrev{cp}{CP}{charge parity}{Product of charge and parity
  transformations}

\newduneabbrev{cpt}{CPT}{charge, parity, and time reversal symmetry}{product of charge, parity
  and time-reversal transformations}

\newduneabbrev{cpv}{CPV}{charge-parity symmetry violation}{Lack of
  symmetry in a system before and after charge and parity
  transformations are applied. 
  For CP symmetry to hold,  a particle turns into its
 corresponding antiparticle under a charge transformation, and a parity
transformation inverts its space coordinates, i.e., 
produces the mirror image}

\newduneword{doe}{DOE}{U.S. Department of Energy}

\newduneabbrev{fra}{FRA}{Fermi Research Alliance}{A joint partnership of the University of Chicago and the Universities Research Association (URA) that manages and operates Fermilab on behalf of the \gls{doe}}

\newduneabbrev{dune}{DUNE}{Deep Underground Neutrino Experiment}{A leading-edge, international experiment for neutrino science and proton decay studies}

\newduneabbrev{esh}{ES\&H}{environment, safety and health}{A discipline and specialty that studies and implements practical aspects of environmental protection and safety at work} 

\newduneabbrev{ppe}{PPE}{personnel protective equipment}{Equipment worn to minimize exposure to hazards that cause serious workplace injuries and illnesses}

\newduneabbrev{odh}{ODH}{oxygen deficiency hazard}{a hazard that occurs when inert gases such as nitrogen, helium, or argon displace room air and thus reduce the percentage of oxygen below the level required for human life}

\newduneabbrev{feshm}{FESHM}{Fermilab Environment, Safety and Health Manual}{The document that contains Fermilab's policies and procedures designed to manage environment, safety, and health in all its programs}

\newduneabbrev{fscf}{FSCF}{far site conventional facilities}{The
  \gls{cf} at the DUNE far detector site, \surf}
  
\newduneabbrev{nscf}{NSCF}{near site conventional facilities}{The
  \gls{cf} at the DUNE near detector site, \fnal}

\newduneabbrevs{gut}{GUT}{grand unified theory}{grand unified theories}{A class of theories that unifies the electro-weak and strong forces}

\newduneabbrev{lar}{LAr}{liquid argon}{Argon in its liquid phase; it is a cryogenic liquid with a boiling point of $\SI{-90}{^\circ{C}}$ (\SI{87}{K}) and density of \SI{1.4}{g/ml}}

\newduneabbrev{lbl}{LBL}{long-baseline}{Refers to the distance between the 
  neutrino source  and the \gls{fd}.  It can also refer to the distance between the near and far detectors. 
  The ``long'' designation is an approximate and relative distinction. For DUNE, this distance  (between \gls{fnal} and \gls{surf}) is approximately \SI{1300}{km}}

\newduneabbrev{lbnf}{LBNF}{Long-Baseline Neutrino Facility}{The
  organizational entity responsible for developing the neutrino beam, the cryostats
  and cryogenics systems, and the conventional facilities for DUNE}
  
\newduneabbrev{lbnf-dune}{LBNF/DUNE}{LBNF and DUNE project}{The overall global project, including \gls{lbnf} and \gls{dune}}

\newduneabbrev{lbnc}{LBNC}{Long-Baseline Neutrino Committee}{The committee, composed of internationally prominent scientists with relevant expertise, charged by the \gls{fnal} director to review the scientific, technical, and managerial progress, plans and decisions associated with \gls{dune}}

\newduneabbrev{ncg}{NCG}{Neutrino Cost Group}{A group of internationally prominent scientists with relevant experience that is charged by the \gls{fnal} director to review the cost, schedule, and associated risks for the \gls{dune} experiment}

\newduneabbrev{mh}{MH}{mass hierarchy}{Describes the separation
  between the mass squared differences related to the solar and
  atmospheric neutrino problems}

\newduneabbrev{mi}{MI}{Fermilab Main Injector}{An accelerator at
  \fnal that provides a beam of high-energy protons that upon
  striking a target produce secondaries that decay to provide the
  neutrinos directed toward the DUNE far detector}

\newduneabbrev{pot}{POT}{protons on target}{Typically used as a unit
  of normalization for the number of protons striking the neutrino
  production target}

\newduneabbrev{qa}{QA}{quality assurance}{The set of actions taken to provide confidence that quality requirements are fulfilled, and to detect and correct poor results}

\newduneabbrev{qc}{QC}{quality control}{An aggregate of activities (such as design analysis and inspection for defects) performed to ensure adequate quality in manufactured products}

\newduneabbrev{sm}{SM}{standard model}{Refers to a theory describing
  the interaction of elementary particles}

\newduneabbrev{tdr}{TDR}{technical design report}{Formal project
  documents 
  that describe the experiment at a technical level, References \cite{Abi:2020wmh, Abi:2020evt, Abi:2020oxb, Abi:2020loh}.}
  
\newduneabbrev{prelimdr}{PDR}{preliminary design report}{A formal project
  document 
  that describes the experiment at a preliminary design level}

\newduneabbrev{tp}{IDR}{interim design report}{An intermediate
milestone on the path to a full \gls{tdr}} 

\newduneabbrev{ckm}{CKM matrix}{Cabibbo-Kobayashi-Maskawa
  matrix}{Refers to the matrix describing the mixing between mass and
  weak eigenstates of quarks}

\newduneabbrev{cl}{CL}{confidence level}{Refers to a probability
  used to determine the value of a random variable given its
  distribution}

\newduneabbrev{pmns}{PMNS}{Pontecorvo-Maki-Nakagawa-Sakata}{A type of matrix that describes the mixing between mass and weak eigenstates of
  the neutrino}

\newduneabbrevs{cpa}{CPA}{cathode plane assembly}{cathode plane assemblies}{The component of the \single detector module that provides the drift HV cathode}

\newduneabbrev{fc}{FC}{field cage}{The component of a \gls{lartpc} that contains and shapes the applied \efield}

\newduneword{cpafc}{CPA/FC}{A pair of \gls{cpa} panels and the top and bottom \gls{fc} portions that attach to the pair; an intermediate assembly for installation into the \gls{spmod} }

\newduneabbrev{topfc}{top FC}{top field cage}{The horizontal portions of the \gls{sp} \gls{fc}   on the top of the \gls{tpc}}

\newduneabbrev{botfc}{bottom FC}{bottom field cage}{The horizontal portions of the \gls{sp} \gls{fc} on the bottom of the \gls{tpc}}

\newduneabbrev{ewfc}{endwall FC}{endwall field cage}{The vertical portions of the \gls{sp} \gls{fc} near the wall}

\newduneabbrev{gp}{GP}{ground plane}{An electrode held electrically neutral relative to Earth ground voltage; it is mounted on the \gls{fc} in a \gls{spmod} to protect the cryostat wall}

  \newduneword{gg}{ground grid}{An electrode held electrically neutral relative to Earth ground voltage; it is installed between the cathode and the \glspl{pd} in a \gls{dpmod} to protect the \glspl{pmt}, maintaining high transparency to light}

\newduneabbrev{alara}{ALARA}{as low as reasonably
  achievable}{Typically used with regard management of radiation
  exposure but may be used more generally. It means making every
  reasonable effort to maintain e.g., exposures, to as far below the
  limits as practical, consistent with the purpose for that the
  activity is undertaken}

\newduneabbrev{ecal}{ECAL}{electromagnetic calorimeter}{A detector
  component that measures energy deposition of traversing particles (in the near detector conceptual design)}

\newduneabbrev{hv}{HV}{high voltage}{Generally describes a voltage
  applied to drive the motion of free electrons through some media, e.g., LAr}

\newduneword{spmod}{SP module}{single-phase DUNE \gls{fd} module}
\newduneword{dpmod}{DP module}{dual-phase DUNE \gls{fd} module}

\newduneabbrev{tcoord}{TC}{technical coordinator}{A member of the \gls{dune} management team responsible for organizing the technical aspects of the project effort; is head of \gls{tc}}

\newduneabbrev{rcoord}{RC}{resource coordinator}{A member of the \gls{dune} management team responsible for coordinating the financial resources of the project effort}

\newduneword{tc}{technical coordination}{The DUNE organization responsible for overall integration 
of the detector elements and successful execution of the detector
construction project; areas of responsibility include 
general project oversight, systems engineering, \gls{qa} 
and safety}

\newduneabbrev{exb}{EB}{executive board}{The highest level DUNE
  decision-making body for the collaboration}

\newduneabbrev{tb}{TB}{technical board}{The DUNE organization responsible for
  evaluating technical decisions}

\newduneabbrev{rrb}{RRB}{Resources Review Board}{A part of \gls{dune}'s international project governance structure, composed of representatives of all funding agencies that sponsor the project, and of  \gls{fnal} management, established to provide coordination among funding partners and oversight of \gls{dune}}

\newduneabbrev{inc}{INC}{International Neutrino Council}{A highest-level international advisory body to the U.S. \gls{doe} and the  \gls{fnal} directorate on matters related to the  \gls{lbnf} and the  \gls{pip2} projects. This council is composed of representatives from the international funding agencies and  \gls{cern} that make major contributions the infrastructure}

\newduneabbrev{cc}{CC}{charged current}{Refers to an interaction
  between elementary particles where a charged weak force carrier
  ($W^+$ or $W^-$) is exchanged}

\newduneabbrev{dis}{DIS}{deep inelastic scattering}{Refers to the 
  interaction of an elementary charged particle with a nucleus in an
  energy range where the interaction can be modeled as taking place with
  individual nucleons}

\newduneabbrev{fsi}{FSI}{final-state interactions}{Refers to
  interactions between elementary or composite particles subsequent to
  the initial, fundamental particle interaction, such as may occur as
  the products exit a nucleus}

\newduneword{geant4}{Geant4}{A
  software toolkit for the simulation of the passage of particles
  through matter using \gls{mc} methods}

\newduneabbrev{genie}{GENIE}{Generates Events for Neutrino Interaction
  Experiments}{Software providing an object-oriented neutrino
  interaction simulation resulting in kinematics of the products of
  the interaction}

\newduneabbrev{mc}{MC}{Monte Carlo}{Refers to a method of numerical
  integration that entails the statistical sampling of the integrand
  function. 
  Forms the basis for some types of detector and physics simulations}

\newduneabbrev{qe}{QE}{quasi-elastic}{Refers to interaction between
  elementary particles and a nucleus in an energy range where the
  interaction can be modeled as occurring between constituent quarks
  of one nucleon and resulting in no bulk recoil of the resulting
  nucleus}

\newduneabbrev{mou}{MoU}{memorandum of understanding}{A document
  summarizing an agreement between two or more parties}

\newduneabbrev{pip2}{PIP-II}{Proton Improvement Plan II}{A \gls{fnal} project for
  improving the protons on target delivered delivered by the \gls{lbnf} neutrino production beam. 
  This is version two of this plan and it is planned to be followed by a PIP-III}
  
\newduneabbrev{sdsta}{SDSTA}{South Dakota Science and Technology
  Authority}{The legal entity that manages \gls{surf}, in Lead, S.D}
  
\newduneabbrev{sdsd}{SDSD}{Fermilab South Dakota Services Division}{A Fermilab division responsible providing host laboratory functions at SURF in South Dakota}

\newduneabbrev{firus}{FIRUS}{Facility Information Reporting Utility System}
 {The safety system at \surf}

\newduneabbrev{bsi}{BSI}{building and site infrastructure}
 {The work package for outfitting of the \gls{lbnf} underground infrastructure}

\newduneabbrev{wbs}{WBS}{work breakdown structure}{An organizational
  project management tool by which the tasks to be performed are
  partitioned in a hierarchical manner}

\newduneabbrev{br}{BR}{branching ratio}{A fractional probability for a
  decay of a composite particle to occur into some specified set or
  sets of products}
\newduneword{bsm}{BSM}{beyond the standard model}

\newduneabbrev{dm}{DM}{dark matter}{The term given to the unknown
  matter or force that explains measurements of galaxy motion 
  that are otherwise inconsistent with the amount of mass associated
  with the observed amount of photon production}
  
  \newduneabbrev{bdm}{BDM}{boosted dark matter}{A new model that describes a relativistic dark matter particle boosted by the annihilation of heavier dark matter participles in the galactic center or the sun}

\newduneabbrev{cern}{CERN}{European Organization for Nuclear
Research}{The leading particle physics laboratory in Europe and home to the ProtoDUNEs. (In French, the Organisation Europ\'{e}enne pour la Recherche Nucl\'{e}aire, derived from Conseil Europ\'{e}en pour la Recherche Nucl\'{e}aire}

\newduneabbrev{dsnb}{DSNB}{diffuse supernova neutrino background}{The
  term describing the pervasive, constant flux of neutrinos due to all
  past supernova neutrino bursts}

\newduneabbrev{espp}{ESPP}{European Strategy for Particle Physics}{The
cornerstone of Europe's
decision-making process for the long-term future of the
field. Mandated by the \gls{cern} Council, it is formed through a broad
consultation of the grass-roots particle physics community, it
actively solicits the opinions of physicists from around the world,
and it is developed in close coordination with similar processes in
the USA and Japan in order to ensure coordination between regions and
optimal use of resources globally}

\newduneabbrev{gar}{GAr}{gaseous argon}{argon in its gas phase}
\newduneabbrev{gartpc}{GArTPC}{gaseous argon time-projection chamber}{A \gls{tpc} filled with gaseous argon; a possible technology choice for the \gls{nd}}

\newduneabbrev{globes}{GLoBES}{General Long-Baseline Experiment
  Simulator}{A software package for simulating energy spectra of
  neutrino flux, interactions, and energy spectra measured after application of some
  model of a detector response)}

\newduneabbrev{snowglobes}{SNOwGLoBES}{SuperNova
Observatories with GLoBES} {From the official description~\cite{snowglobes}: 
SNOwGLoBES is public software for computing interaction rates and distributions of observed quantities for \gls{snb} neutrinos in common detector materials} 

\newduneword{l/e}{L/E}{length-to-energy ratio}
\newduneword{lri}{LRI}{long-range interactions}

\newduneabbrev{nc}{NC}{neutral current}{Refers to an interaction
  between elementary particles where a neutrally charged weak force carrier
  ($Z^0$) is exchanged}

\newduneabbrev{nh}{NH}{normal hierarchy}{Refers to the neutrino mass
  eigenstate ordering whereby the sign of the mass squared difference
  associated with the atmospheric neutrino problem is positive}

\newduneabbrev{ih}{IH}{inverted hierarchy}{Refers to the neutrino mass
  eigenstate ordering whereby the sign of the mass squared difference
  associated with the atmospheric neutrino problem is negative}

\newduneabbrev{no}{NO}{normal ordering}{Refers to the neutrino mass
  eigenstate ordering whereby the sign of the mass squared difference
  associated with the atmospheric neutrino problem is positive}

\newduneabbrev{io}{IO}{inverted ordering}{Refers to the neutrino mass
  eigenstate ordering whereby the sign of the mass squared difference
  associated with the atmospheric neutrino problem is negative}

\newduneabbrev{msw}{MSW}{Mikheyev-Smirnov-Wolfenstein effect}{Explains
  the oscillatory behavior of neutrinos produced inside the sun as
  they traverse the solar matter}

\newduneabbrev{nsi}{NSI}{nonstandard interaction}{A general class of
  theory of elementary particles other than the Standard Model}

\newduneabbrev{pfive}{P5}{Particle Physics Project Prioritization
Panel}{The Particle Physics Project Prioritization Panel (P5) was a
subpanel of the High Energy Physics Advisory Panel (HEPAP). It completed
its Report, a ten-year strategic plan for high energy physics in the
U.S., in 2014. This report included a recommendation that ``host a world-leading neutrino
program that will have an optimized set of short- and long-baseline neutrino oscillation experiments, and its long-term focus
is a reformulated venture referred to here as the Long Baseline
Neutrino Facility (LBNF)''}

\newduneabbrev{sme}{SME}{standard-model extension}{an effective field theory that contains the \gls{sm}, general relativity, and all possible operators that break Lorentz symmetry (Wikipedia)}

\newduneabbrev{susy}{SUSY}{supersymmetry}{Theoretical symmetry between a fermion and a boson}

\newduneabbrev{wimp}{WIMP}{weakly-interacting massive particle}{A
  hypothesized particle that may be a component of dark matter}

\newduneabbrev{ce}{CE}{cold electronics}{Analog and digital readout electronics that operate at cryogenic temperatures}

\newduneabbrev{crp}{CRP}{charge-readout plane}{In the \gls{dp} technology, a  collection of
  electrodes in a planar arrangement placed at a particular voltage
  relative to some applied \efield such that drifting electrons
  may be collected and their number and time may be measured}

\newduneabbrev{dram}{DRAM}{dynamic random access memory}{A computer memory technology}

\newduneabbrev{fnal}{Fermilab}
{Fermi National Accelerator Laboratory}{U.S. national laboratory in Batavia, IL. It is the laboratory that hosts \gls{dune} and serves as its near site}

\newduneabbrev{bnl}{BNL}{Brookhaven National Laboratory}{US national laboratory in Upton, NY}

\newduneabbrev{slac}{SLAC}{SLAC National Accelerator Laboratory}{US national laboratory in Menlo Park, CA}

\newduneabbrev{lbnl}{LBNL}{Lawrence Berkeley National Laboratory}{US national laboratory in Berkeley, CA}

\newduneabbrev{anl}{ANL}{Argonne National Laboratory}{US national laboratory in Lemont, IL}

\newduneabbrev{lanl}{LANL}{Los Alamos National Laboratory}{US national laboratory in Los Alamos, NM}

\newduneabbrev{fs}{FS}{full stream}{Relates to a data stream that has not undergone selection, compression or other form of reduction}

\newduneabbrev{lem}{LEM}{large electron multiplier}{A micro-pattern detector suitable for use in ultra-pure argon vapor; LEMs consist of copper-clad PCB boards with sub-millimeter-size holes through which electrons undergo amplification}

\newduneabbrev{lng}{LNG}{liquefied natural gas}{Pertaining to natural gas in its liquid phase}

\newduneabbrev{mip}{MIP}{minimum ionizing particle}{Refers to a
  particle traversing some medium such that the particle's mean energy loss is  
  near the minimum}

\newduneabbrev{pd}{PD}{photon detector}{The detector
  elements involved in measurement of the number and arrival times of
  optical photons produced in a detector module} 

\newduneabbrev{pmt}{PMT}{photomultiplier tube}{A device that makes use
  of the photoelectric effect to produce an electrical signal from the
  arrival of optical photons}

\newduneabbrev{ppm}{ppm}{parts per million}{A concentration equal to one part in $10^{-6}$}
\newduneabbrev{ppb}{ppb}{parts per billion}{A concentration equal to one part in $10^{-9}$}
\newduneabbrev{ppt}{ppt}{parts per trillion}{A concentration equal to one part in $10^{-12}$}

\newduneword{rio}{RIO}{reconfigurable input output}

\newduneabbrev{s/n}{S/N}{signal-to-noise}{signal-to-noise ratio}
\newduneword{snr}{\mbox{S/N}}{signal-to-noise ratio}

\newduneword{ssp}{SSP}{SiPM signal processor}

\newduneabbrev{sbn}{SBN}{Short-Baseline Neutrino}{A \gls{fnal} program consisting of three collaborations, \gls{microboone}, \gls{sbnd}, and \gls{icarus}, to perform sensitive searches for $\nue$ appearance and $\numu$ disappearance in the Booster Neutrino Beam}

\newduneword{stt}{STT}{straw tube tracker}

\newduneword{wire board}{wire board}{At the head end of the APA in the \single TPC, stacks of electronics boards referred to as ``wire boards'' are arrayed to anchor the wires.  They also provide the connection between the wires and the cold electronics} 

\newduneabbrev{wls}{WLS}{wavelength-shifting}{A material or process by
  which incident photons are absorbed by a material and photons are
  emitted at a different, typically longer, wavelength}
  
\newduneabbrev{tpb}{TPB}{tetra-phenyl butadiene}{A 
\gls{wls} material}

\newduneabbrev{ptp}{PTP}{p-terphenyl}{A 
\gls{wls} material}

\newduneabbrev{sft}{SFT}{signal feedthrough}{A cryostat penetration allowing for the passage of cables or other extended parts}
\newduneabbrev{sftchimney}{SFT chimney}{signal feedthrough chimney}{In the \dual technology, a volume above the cryostat penetration used for a signal feedthrough}

\newduneabbrev{catiroc}{CATIROC}{charge and time integrated readout chip}{A complete read-out chip manufactured in AustriaMicroSystem designed to read arrays of 16 photomultipliers}

\newduneabbrev{wr}{WR}{White Rabbit}{A component of the timing system that forwards clock signal and time-of-day reference data to the master timing unit}

\newduneabbrev{mch}{MCH}{MicroTCA Carrier Hub}{An network switching device}

\newduneabbrev{wrmch}{WR-MCH}{White Rabbit \gls{utca} Carrier Hub}{A card mounted in \gls{utca} crate that recieves time syncronization information and trigger data packets over \gls{wr} network and disributes them to the \gls{amc} over \gls{utca} backplane} 

\newduneabbrev{wrtsn}{WR-TSN}{White Rabbit TimeStamping Node}{A unit on the \gls{wr} network that timestamps the trigger signals and sends out trigger data packets to \gls{wrmch}}

\newduneword{cmp}{CMP}{configuration management plan}
\newduneword{qap}{QAP}{quality assurance plan} 
\newduneword{ieshp}{IESHP}{integrated environmental, safety and health plan}
\newduneword{dmp}{DMP}{data management plan} 
\newduneword{qam}{QAM}{quality assurance manager} 

\newduneabbrev{dss}{DSS}{detector support system}{The system used to support a \gls{sp} \gls{detmodule} within its cryostat}

\newduneabbrev{ddss}{DDSS}{DUNE detector safety system}{The system used to manage key aspects of detector safety}

\newduneabbrev{lc}{LC}{logistics center}{A facility where \gls{lbnf} and \gls{dune} components will be received and transhipped to \gls{surf}}

\newduneabbrev{tco}{TCO}{temporary construction opening}{An opening in the side of a cryostat through which detector elements are brought into the cryostat; utilized during construction and installation}

\newduneabbrev{surf}{SURF}{Sanford Underground Research Facility}{The laboratory in South Dakota where the \gls{lbnf} \gls{fscf} will be constructed and the \gls{dune} \gls{fd} will be installed and operated}

\newduneabbrev{sit}{SIT}{surface installation team}{An organizational unit responsible for logistics and integration in South Dakota}

\newduneabbrev{uit}{UIT}{underground installation team}{An organizational unit responsible for installation in the underground area at the \gls{surf} site}

\newduneabbrev{cmgc}{CMGC}{construction manager/general contractor}{The organizational unit responsible for management of the construction of conventional facilities at the underground area at the \surf site}

\newduneword{cdrev}{conceptual design review}{A project management device by which a conceptual design is reviewed} 
\newduneword{pdr}{preliminary design review}{A project management device by which an early design is reviewed} 
\newduneword{fdr}{final design review}{A project management device by which a final design is reviewed}
\newduneword{prr}{production readiness review}{A project management device by which the production readiness is reviewed}
\newduneword{irr}{installation readiness review}{A project management device by which the plan for installation is reviewed}
\newduneword{orr}{operational readiness review}{A project management device by which the operational readiness is reviewed}
\newduneword{ppr}{production progress review}{A project management device by which the progress of production is reviewed} 
\newduneabbrev{edms}{EDMS}{engineering document management system}{A computerized document management system developed and supported at \gls{cern} in which some DUNE documents, drawings and engineering models are managed}
\newduneabbrev{ecr}{ECR}{engineering change request}{The first step in the change control process in which a proposed change is described}
\newduneabbrev{docdb}{DocDB}{Document DataBase}{A computerized document management system developed and supported at \gls{fnal} in which virtually all LBNF and most DUNE documents are managed}

\newduneword{wrgm}{WR grandmaster}{White Rabbit grandmaster}

\newduneabbrev{larsoft}{LArSoft}{Liquid Argon Software}{A shared base of physics software across \lartpc experiments}
\newduneword{nova}{NOvA}{The \nova off-axis neutrino oscillation experiment at \gls{fnal}}
\newduneword{minerva}{MINERvA}{The \minerva neutrino cross sections experiment at  \gls{fnal}}
\newduneword{microboone}{MicroBooNE}{The \lartpc-based \microboone neutrino oscillation experiment at  \gls{fnal}}
\newduneword{sbnd}{SBND}{The Short-Baseline Near Detector experiment at  \gls{fnal}}
\newduneabbrev{nexo}{nEXO}{Enriched Xenon Observatory}{Experiment at Lawrence Livermore National Laboratory (U.S. national lab in Livermore, CA)searching for new physics with neutrinoless double-beta decay}
\newduneword{argoneut}{ArgoNeuT}{The ArgoNeuT test-beam experiment and \gls{lartpc} prototype at  \gls{fnal}}
\newduneword{icarus}{ICARUS}{A neutrino experiment that was located at the Laboratori Nazionali del Gran Sasso (LNGS) in Italy, then refurbished at \gls{cern} for re-use in the same neutrino beam from \gls{fnal} used by the MiniBooNE, \gls{microboone} and \gls{sbnd} experiments. The ICARUS detector is being reassembled at \gls{fnal}}
\newduneword{atlas}{ATLAS}{One of two general-purpose detectors at the \gls{lhc}. It investigates a wide range of physics, from the search for the Higgs boson to extra dimensions and particles that could make up \gls{dm}}

\newduneword{lbne}{LBNE}{Long Baseline Neutrino Experiment (a terminated US project that was reformulated in 2014 under the auspices of the new \gls{dune} collaboration, an internationally coordinated and internationally funded program, with \gls{fnal} as host)}

\newduneabbrev{lbno}{LBNO}{Long Baseline Neutrino Observatory} {A terminated European project that, during its six-year duration, assessed the feasibility of a next-generation deep underground neutrino observatory in Europe)}

\newduneword{wirecell}{Wire-Cell}{A tomographic automated \threed neutrino event reconstruction method for \lartpc{}s}
\newduneabbrev{wct}{WCT}{Wire-Cell Toolkit}{A software toolkit with data flow processing components for \lartpc noise and signal simulation, noise filtering, signal processing, and tomographic \threed ionization activity imaging}
\newduneword{ftslite}{F-FTS-lite}{Light-weight version of the \fnal File Transfer system used for rapid data transfers out of the online systems}
\newduneabbrev{fts}{FTS}{File Transfer System}{A file transfer system developed at \fnal to catalog and move data to permanent storage}

\newduneword{35t}{35 ton prototype}{A prototype cryostat and \gls{sp} detector built at \fnal before the \gls{protodune} detectors}

\newduneabbrev{mcr}{MCR}{main communications room}{Space at the \gls{fd} site for cyber infrastructure}

\newduneabbrev{cuc}{CUC}{central utility cavern}{The utility cavern at the 4850L of \gls{surf} located between the two detector caverns. It contains utilities such as central cryogenics and other systems, and the underground data center and control room}

\newduneabbrev{cfd}{CFD}{computational fluid dynamics}{High performance computer-assisted modeling of fluid dynamical systems}
\newduneword{vuv}{VUV}{vacuum ultra-violet}
\newduneword{tallbo}{TallBo}{A cylindrical cryostat at \gls{fnal} primarily used for developing scintillation light collection technologies for \gls{lartpc} detectors}

\newduneword{root}{ROOT}{A modular scientific software toolkit. It provides all the functionalities needed to deal with big data processing, statistical analysis, visualisation and storage. It is mainly written in C++ but integrated with other languages such as Python and R}

\newduneabbrev{eos}{EOS}{EOS}{The XRootD-based distributed file system developed by CERN}
\newduneabbrev{ehn1}{EHN1}{Experiment Hall North One}{Location at CERN of the ProtoDUNE experiments}
\newduneword{led}{LED}{Light-emitting diode}
\newduneabbrev{rtd}{RTD}{resistance temperature detector}{A temperature sensor consisting of a material with an accurate and reproducible resistance/temperature relationship}
\newduneword{swc}{SWC}{Software \& Computing}
\newduneabbrev{las}{LAS}{LEM-anode Sandwich}{In the \dual technology, a \gls{lem} and its corresponding anode are mounted together in a module called a LEM-anode sandwich}

\newduneword{roi}{ROI}{region of interest}
\newduneabbrev{hpc}{HPC}{high-performance computing}{high-performance computing facilities; generally computing facilities emphasizing parallel computing with aggregate power of more than a teraflop}

\newduneword{comfund}{common fund}{The shared resources of the collaboration}
\newduneabbrev{ims}{IMS}{integrated master schedule}{A project management device consisting of linked tasks and milestones}

\newduneword{hvdb}{HVDB}{HV divider board}
\newduneword{sas}{SAS}{Another term for the materials airlock; a pass-through chamber used to ensure safe transfer of materials into a clean room, avoiding contamination in both directions}

\newduneabbrev{fea}{FEA}{finite element analysis}{Simulation of a physical phenomenon using the numerical technique called Finite Element Method (FEM), a numerical method for solving problems of engineering and mathematical physics}

\newduneword{fss}{FSS}{field shaping strips}
\newduneword{lvds}{LVDS}{low-voltage differential signaling}

\newduneword{esd}{ESD}{electrostatic discharge}

\newduneabbrev{rp}{RP}{resistive panel}{Resistive panels form the constant potential surfaces for a \gls{spmod} \gls{cpa}; they are composed of a thin layer of carbon-impregnated Kapton and laminated to both sides of a \frfour sheet}

\newduneword{uhmwpe}{UHMWPE}{ultra-high molecular weight polyethylene}

\newduneword{cts}{CTS}{Cryogenic Test System}
\newduneword{plc}{PLC}{programmable logic controller}

\newduneword{mppc}{MPPC}{\SI{6}{mm}$\times$\SI{6}{mm} Multi-Pixel Photon Counters produced by Hamamatsu\texttrademark{} Photonics K.K}

\newduneabbrev{sfp}{SFP}{small form-factor pluggable}{a particular standard for optical transceivers}

\newduneabbrev{minipod}{MiniPOD}{miniature parallel optical device}{a family of types of multi-channel optical transceivers}

\newduneword{ccc}{CCC}{configuration change command}
\newduneword{act}{ACT}{activation time stamp}
\newduneword{lcm}{LCM}{light calibration module}
\newduneword{lpm}{LPM}{light pulser module}
\newduneword{dac}{DAC}{digital-to-analog converter}
\newduneword{arapuca}{ARAPUCA}{A \gls{pds} design that consists of a light trap that captures wavelength-shifted photons inside boxes with highly reflective internal surfaces until they are eventually detected by \gls{sipm} detectors or are lost}
\newduneword{sarapu}{S-ARAPUCA}{Standard \gls{arapuca} design with different \gls{wls} coatings on both faces of the dichroic filter window(s) of the cell}
\newduneword{xarapu}{X-ARAPUCA}{Extended \gls{arapuca} design with \gls{wls} coating on only the external face of the dichroic filter window(s) but with a \gls{wls} doped plate inside the cell}
\newduneword{feb}{FEB}{front-end board}

\newduneabbrev{lsnd}{LSND}{Liquid Scintilator Neutrino Detector}{A scintillation detector and associated experiment located at Los Alamos National Laboratory}

\newduneabbrev{cvn}{CVN}{convolutional visual network}{An algorithm for identifying neutrino interactions based on their topology and without the need for detailed reconstruction algorithms}

\newduneword{pandora}{Pandora}{The Pandora multi-algorithm approach to pattern recognition} 

\newduneabbrev{pma}{PMA}{Projection Matching Algorithm}{A reconstruction algorithm that combines \twod reconstructed objects to form a \threed representation}
\newduneabbrev{bdt}{BDT}{boosted decision tree}{A method of multivariate analysis}
\newduneabbrev{cnn}{CNN}{convolutional neural network}{A deep learning technique most commonly applied to analyzing visual imagery}
\newduneword{pdg}{PDG}{Particle Data Group}

\newduneword{pci}{PCI}{Peripheral Component Interconnect}

\newduneword{labview}{LabVIEW}{Laboratory Virtual Instrument Engineering Workbench is a system-design platform and development environment for a visual programming language from National Instruments}

\newduneword{pcb}{PCB}{printed circuit board}

\newduneword{crio}{cRIO}{Compact Reconfigurable Input Output}

\newduneword{dcs}{DCS}{Distributed Communications System}

\newduneword{opc-ua}{OPC-UA}{OPC  Unified Architecture is a machine to machine communication protocol for industrial automation developed by the OPC Foundation. OPC stands for Object Linking and Embedding for Process Control}

\newduneword{cabangle}{Cabibbo angle}{A quark mixing parameter that governs the coupling of up quarks to strange quarks}
\newduneword{valor}{VALOR}{A neutrino oscillation fitting framework that is used by \gls{t2k}; the name stands for VALencia-Oxford-Rutherford, the original three institutions that developed it}
\newduneword{cafana}{CAFAna}{Common Analysis File Analysis}
\newduneabbrev{pca}{PCA}{principal component analysis}{A statistical procedure that uses an orthogonal transformation to convert a set of observations of possibly correlated variables into a set of values of linearly uncorrelated variables called principal components (Wikipedia)}
\newduneword{numi}{NuMI}{a set of facilities at \fnal, collectively called ``Neutrinos at the Main Injector.''  The NuMI neutrino beamline target system converts an intense proton beam into a focused neutrino beam}
\newduneword{gibuu}{GiBUU}{Giessen Boltzmann-Uehling-Uhlenback Project; a unified theory and transport framework in the MeV and GeV energy regimes for elementary reactions on nuclei }
\newduneabbrev{rpa}{RPA}{random phase approximation} {an approximation method commonly used for describing the dynamic linear electronic response of electron systems (Wikipedia)}
\newduneword{t2k}{T2K}{T2K (Tokai to Kamioka) is a long-baseline neutrino experiment in Japan studying neutrino oscillations}
\newduneword{mptdet}{MPT detector}{multipurpose tracking detector}

\newduneword{lariat}{LArIAT}{The repurposed ArgoNeuT \gls{lartpc}, modified for use in a charged particle beam, dedicated to the calibration and precise characterization of the output response of these detectors}

\newduneword{captain}{CAPTAIN}{Experimental program sited at \gls{lanl} that is designed to make measurements of scientific importance to \gls{lbl} neutrino physics and physics topics that will be explored by large underground detectors}

\newduneword{dayabay}{Daya Bay}{a neutrino-oscillation experiment in Daya Bay, China, designed to measure the mixing angle $\Theta_{13}$  using antineutrinos produced by the reactors of the Daya Bay and Ling Ao nuclear power plants}

\newduneword{nuwro}{NuWro}{neutrino interaction generator}

\newduneabbrev{neut}{NEUT}{neutrino interaction generator}{A neutrino interaction simulation program library for the studies of atmospheric accelerator neutrinos}

\newduneword{minos}{MINOS}{A long-baseline neutrino experiment, with a near detector at \gls{fnal} and a far detector in the Soudan mine in Minnesota, designed to observe the phenomena of neutrino oscillations (ended data runs in 2012)}

\newduneabbrev{efig}{EFIG}{Experimental Facilities Interface Group}{The body responsible for the required high-level coordination between the \gls{lbnf} and \gls{dune} projects}
\newduneword{ashriver}{Ash River}{The Ash River, Minnesota, USA \gls{nova} experiment far site, used as an assembly test site for \gls{dune}} 

\newduneword{ipd}{project integration director}{Responsible for integration and installation of \gls{lbnf} and \gls{dune} deliverables in South Dakota. Manages the \gls{integoff}}

\newduneabbrev{jpo}{JPO}{Joint Project Office}{The framework through which team members from the LBNF project office, \gls{integoff}, and DUNE \gls{tc} work together to provide coherence in project support functions across the global enterprise. 
Its functions include global project configuration and integration, installation planning and coordination, scheduling, safety assurance, technical review planning and oversight, development of partner agreements, and financial reporting}

\newduneword{ifbeam}{IFbeam}{Database that stores beamline information 
indexed by timestamp}

\newduneabbrev{marley}{MARLEY}{Model of Argon Reaction Low Energy
Yields}{Developed at UC Davis, MARLEY is the first realistic model of
neutrino electron interactions on argon for enegies less than \SI{50}{MeV}. This includes the energy range important for \gls{snb}
neutrinos and also solar 8--boron neutrinos}

\newduneabbrev{es}{ES}{elastic scattering}{Events in which a neutrino
elastically scatters off of another particle}

\newduneabbrev{cno}{CNO}{carbon nitrogen oxygen}{The CNO cycle (for carbon-nitrogen-oxygen) is one of the two known sets of fusion reactions by which stars convert
hydrogen to helium, the other being the proton-proton chain reaction
(pp-chain reaction). In the CNO cycle, four protons fuse, using
carbon, nitrogen, and oxygen isotopes as catalysts, to produce one
alpha particle, two positrons and two electron neutrinos}

\newduneabbrev{sdwf}{SDWF}{South Dakota Warehouse Facility}{Warehousing operations in South Dakota responsible for receiving LBNF and DUNE goods and coordinating shipments to the Ross shaft at \gls{surf}}

\newduneabbrev{wms}{WMS}{warehouse management system}{Commercial software package used to track shipments and interface to freight forwarders. This includes a database for shipping}

\newduneabbrev{dcdb}{DCDB}{DUNE construction database}{Database used by DUNE to track the history and testing of all parts of each \gls{detmodule}}

\newduneabbrev{aup}{AUP}{acceptance for use and possession}{Required for beneficial occupancy of the underground areas at SURF for LBNF and DUNE}

\newduneabbrev{bms}{BMS}{building management system}{Part of the safety system at \gls{surf} that includes the fire and life safety system}
\newduneabbrev{fls}{FLS}{fire and life safety system}{Part of the safety system at \gls{surf}}

\newduneabbrev{sno}{SNO}{Sudbury Neutrino Observatory}{The Sudbury
Neutrino Observatory was a detector built 6800 feet under ground, in
INCO's Creighton mine near Sudbury, Ontario, Canada. SNO was a
heavy-water Cherenkov detector designed to detect neutrinos produced
by fusion reactions in the sun}

\newduneword{sk}{Super-Kamiokande}{Experiment sited in the Kamioka-mine, Hida-city, Gifu, Japan that uses a large water Cherenkov detector to study neutrino properties through the observation of solar neutrinos, atmospheric neutrinos and man-made neutrinos}

\newduneabbrev{id}{ID}{inner diameter}{Inner diameter of a tube}

\newduneabbrev{od}{OD}{outer diameter}{Outer diameter of a tube}

\newduneabbrev{rms}{RMS}{root mean square}{The square root of the arithmetic mean of the squares of a set of values, used as a measure of the typical magnitude of a set of numbers, regardless of their sign}

\newduneabbrev{orc}{ORC}{operational readiness clearance}{Final safety approval prior to the start of operation}

\newduneabbrev{gsc}{GSC group}{global safety coordination group}{DUNE group that evaluates applicable codes and standards, including international code equivalency, for the design, assembly, and installation of the \gls{fd}}

\newduneabbrev{ha}{HA}{hazard analysis}{A first step in a process to assess risk; the result of hazard analysis is the identification of the hazards present for a task or process}
\newduneword{har}{HAR}{hazard analysis report}

\newduneabbrev{tap}{TAP}{trip action plan}{A document required for any trip by a worker to the underground area at \gls{surf}, per that site's access control program; 
it describes the work to be accomplished during the trip} 

\newduneword{em}{EM}{emergency management}
\newduneword{ert}{ERT}{emergency response team}

\newduneabbrev{ndk}{NDK}{nucleon decay}{The hypothetical, baryon number violating decay of a proton or a bound neutron into lighter particles}

\newduneabbrev{emi}{EMI}{electromagnetic interference}{Disturbance generated by an external source that affects an electrical circuit by electromagnetic induction, electrostatic coupling, or conduction}

\newduneabbrev{pe}{PE}{photoelectron}{An electron ejected from the surface of a material by the photoelectric effect}

\newduneabbrev{spe}{SPE}{single photoelectron}{A single photoelectron}

\newduneabbrev{fwhm}{FWHM}{full width at half maximum}{Width of a distribution measured between those points at which the distribution is equal to half of its maximum amplitude}

\newduneabbrev{gdml}{GDML}{geometry description markup language}{An application-indepedent, geometry-description format based on XML}

\newduneabbrev{xml}{XML}{extensible markup language}{A markup language that defines a set of rules for encoding documents in a format that is both human-readable and machine-readable}

\newduneabbrev{crt}{CRT}{cosmic ray tagger}{Detector external to the TPC designed to tag TPC-traversing cosmic ray particles}

\newduneabbrev{sn}{SN}{supernova}{Event that occurs upon the death of certain types of stars}

\newduneabbrev{wg}{WG}{working group}{A group of persons working together to achieve specified goals}

\newduneabbrev{ctsf}{CTSF}{coating, testing and storage facility}{A facility where the the \dual photon detectors will be coated, tested, and stored}

\newduneword{rucio}{Rucio}{Data management system originally developed
by \gls{atlas} but now open-source and shared across HEP}
\newduneabbrev{doma}{DOMA}{data organization, management, and
access}{data organization, management, and access efforts through the
HEP Software Foundation}

\newduneabbrev{hsf}{HSC}{HEP Software Foundation Collaboration}{A foundation that facilitates cooperation and common efforts in high energy physics software and computing internationally}

\newduneabbrev{wlcg}{WLCG}{Worldwide LHC Computing Grid}{Worldwide LHC
Computing Grid}
\newduneabbrev{osg}{OSG}{Open Science Grid}{Open Science Grid}
\newduneabbrev{sci}{SCI}{Scientific Computing Infrastructure}{Proposed
extension of the infrastructure component of \gls{wlcg} to other
experiments}
\newduneabbrev{csc}{CSC}{computing and software consortium}{DUNE
computing and software consortium}

\newduneword{dirac}{DIRAC}{Computing workflow management designed for
LHCb and now used by many HEP experiments}

\newduneword{frp}{FRP}{fiber-reinforced plastic}
\newduneabbrev{hdpe}{HDPE}{high-density polyethylene}{High-density polyethylene plastic}
\newduneword{hvps}{HVPS}{\gls{hv} power supply}
\newduneword{aisi}{AISI}{American Iron and Steel Institute}
\newduneword{ific}{IFIC}{Instituto de Fisica Corpuscular (in Valencia, Spain)}
\newduneabbrev{rsds}{RSDS}{radioactive source deployment system}{Proposed calibration system based on the deployment of
radioactive sources inside the \gls{dune} cryostat}
\newduneword{2p2h}{2p2h}{two particle, two hole}
\newduneabbrev{duneprism}{DUNE-PRISM}{\gls{dune} Precision Reaction-Independent Spectrum Measurement}{a mobile near detector that can perform measurements over a range of angles off-axis from the neutrino beam direction in order to sample many different neutrino energy distributions}
\newduneword{arcube}{ArgonCube}{The name of the core part of the \gls{dune} \gls{nd}, a \gls{lartpc}}

\newduneabbrev{citf}{CITF}{cryogenic instrumentation test facility}{A facility at \fnal with small ($<\,\SI{1}{ton}$) to intermediate ($\sim\,\SI{1}{ton}$) volumes of instrumented, purified TPC-grade \lar, used for testing devices intended for use in \gls{dune}}

\newduneabbrev{3dst}{3DST}{3D scintillator tracker}{The core part of the \threed projection scintillator tracker spectrometer in the near detector conceptual design}
\newduneabbrev{3dsts}{3DST-S}{3D scintillator tracker spectrometer}{The \threed projection scintillator tracker spectrometer  in the near detector conceptual design}
\newduneabbrev{mpd}{MPD}{multi-purpose detector}{A component of the near detector conceptual design; it is a magnetized system consisting of a \gls{hpgtpc} and a surrounding \gls{ecal}}
\newduneabbrev{hpg}{HPG}{high-pressure gas}{gas at high pressure to be used in a \gls{hpgtpc}} 
\newduneabbrev{hpgtpc}{HPgTPC}{high-pressure gaseous argon TPC}{A \gls{tpc} filled with gaseous argon; a possible component of the \gls{dune} \gls{nd}}

\newduneword{src}{SRC}{short-range correlated nucleon-nucleon interactions}
\newduneword{larpix}{LArPix}{ \gls{asic} pixelated charge readout for a \gls{tpc} }
\newduneword{arclt}{ArCLight}{a light detector \gls{arcube} effort}
\newduneword{fhc}{FHC}{forward horn current ($\numu$ mode)}
\newduneword{rhc}{RHC}{reverse horn current ($\overline{\nu}_{\mu}$ mode)}
\newduneword{mwpc}{MWPC}{multi-wire proportional chamber}
\newduneword{na61}{NA61}{CERN hadron production experiment}
\newduneword{pdnd}{ProtoDUNE-ND}{a prototype \gls{dune} \gls{nd}}
\newduneword{ccqe}{CCQE}{charged current quasielastic interaction} 
\newduneabbrev{roc}{ROC}{readout chamber}{readout chamber for gaseous argon \gls{tpc}}
\newduneabbrev{iroc}{IROC}{inner readout chamber}{inner (radial) readout chamber for gaseous argon \gls{tpc}}
\newduneabbrev{oroc}{OROC}{outer readout chamber}{outer (radial) readout chamber for gaseous argon \gls{tpc}}

\newduneword{lux}{LUX}{Large Underground Xenon (LUX) dark matter detector at \gls{surf} }

\newduneword{mjdemo}{Majorana Demonstrator}{Experiment sited at \gls{surf} that  seeks to determine whether neutrinos are their own antiparticles}

\newduneword{lz}{LZ}{Experiment sited at \gls{surf} that  seeks to detect faint interactions between galactic dark matter and regular matter}

\newduneword{mu2e}{Mu2e}{An experiment sited at \gls{fnal} that searches for charged-lepton flavor violation and seeks to discover physics beyond the \gls{sm}}

\newduneword{pdsp2}{ProtoDUNE-SP-2}{A second test run in the singe-phase
ProtoDUNE test stand at CERN, acting as a validation of the final
single-phase detector design}

\newduneword{osha}{OSHA}{Occupational Safety and Health Administration (USA Department of Labor) formed by the Occupational Safety and Health Act of 1970}
\newduneabbrev{pns}{PNS}{pulsed neutron source}{Calibration system based
on neutron capture gamma showers spread out in the whole detector}

\newduneabbrev{fv}{FV}{fiducial volume}{The detector volume within the \gls{tpc} 
that is selected for physics analysis through cuts on reconstructed event position}

\newduneword{p6}{P6}{framework used to plan and status the resource-loaded schedule of activities associated with the USA contributions to \gls{lbnf} and \gls{dune} }
\newduneabbrev{evms}{EVMS}{earned value management system}{Earned Value Management is a systematic approach to the integration and measurement of cost, schedule, and technical (scope) accomplishments on a project or task. It provides both the government and contractors the ability to examine detailed schedule information, critical program and technical milestones, and cost data (text from the US DOE); the EVMS is a system that implements this approach}

\newduneword{core}{CORE}{CORE contributions are in either monetary units or labor hours. They can be technical components for the facility or experiment and the effort of the staff needed to produce, install, and test them;  major facilities for the experiment; or other products and services relevant for the completion of the facility or experiment} 

\newduneabbrev{ahj}{AHJ}{Authority Having Jurisdiction}{An organization, office, or individual responsible for enforcing the requirements of a code or standard, or for approving equipment, materials, an installation, or a procedure (OSHA)}
\newduneword{cte}{CTE}{coefficient of thermal expansion}

\newduneabbrev{opc}{OPC}{open platform communications}{Open platform communications is a series of standards and specifications for industrial telecommunication} 
\newduneword{scada}{SCADA}{supervisory control and data acquisition}
\newduneword{ln}{LN$_2$}{liquid nitrogen}
\newduneabbrev{lapd}{LAPD}{Liquid Argon Purity Demonstrator}{Cryostat at Fermilab for long-term studies requiring a large volume of argon}

\newduneabbrev{pab}{PAB}{Proton Assembly Building}{Home of several \gls{lar} facilities at Fermilab}
\newduneword{hep}{HEP}{high energy physics}
\newduneword{sc}{SC}{scientific computing}  
\newduneword{cms}{CMS}{Compact Muon Solenoid experiment at CERN}
\newduneword{alice}{ALICE}{A Large Ion Collider Experiment, at CERN}
\newduneword{gpib}{GPIB}{general purpose interface bus}

\newduneabbrev{pfparticle}{PFParticle}{particle flow particle}{Each of the individual reconstructed particles in the hierarchy (or particle flow) describing the reconstructed event interaction}

\newduneabbrev{mcparticle}{MCParticle}{Monte Carlo Particle}{Individual true simulated particle}
\newduneword{au}{AU}{astronomical unit}
\newduneword{nufit}{NuFIT 4.0}{The NuFIT 4.0 global fit to neutrino oscillation data}

\newduneabbrev{sgft}{SGFT}{term}{add def (DP install)}
\newduneword{uhv}{UHV}{ultra high vacuum}
\newduneword{lps}{LPS}{laser positioning system}

\newduneword{unicamp}{UNICAMP}{University of Campinas, Sao Paulo, Brazil}
 
\newduneabbrev{fbk}{FBK}{Fondazione Bruno Kessler}{FBK is a research non-profit entity in Trento, Italy that partners in the development of technology with applications in various fields including High Energy Physics}

\newduneword{fft}{FFT}{fast Fourier transform}
\newduneabbrev{enob}{ENOB}{effective number of bits}{The effective number of bits is a measure of the dynamic range of an \gls{adc} and its associated circuitry. The resolution of an \gls{adc} is specified by the number of bits used to represent the analog value, in principle giving 2N signal levels for an N-bit signal. However, all real \gls{adc} circuits introduce noise and distortion. ENOB specifies the resolution of an ideal \gls{adc} circuit that would have the same resolution as the circuit under consideration}
\newduneabbrev{sndr}{SNDR}{signal to noise and distortion ratio}{Also known as SINAD. Ratio of the \gls{rms} signal amplitude to the mean value of the root-sum-square of all other spectral components, including harmonics, but excluding \gls{dc} levels. It is a good indication of the overall dynamic performance of an \gls{adc} because it includes all components which make up noise and distortion}
\newduneabbrev{sfdr}{SFDR}{spurious free dynamic range}{Spurious free dynamic range is the ratio of the \gls{rms} value of the signal to the \gls{rms} value of the worst spurious signal regardless of where it falls in the frequency spectrum. The worst spur may or may not be a harmonic of the original signal}
\newduneabbrev{thd}{THD}{total harmonic distortion}{Total harmonic distortion is the ratio of the \gls{rms} value of the fundamental signal to the mean value of the root-sum-square of its harmonics} 
\newduneword{tvs}{TVS}{transient voltage suppression}

\newduneword{riskprob}{risk probabilities}{The risk probability, after taking into account the planned mitigation activities, is ranked as 
L (low $<\,$\SI{10}{\%}), 
M (medium \SIrange{10}{25}{\%}), or 
H (high $>\,$\SI{25}{\%}). 
The cost and schedule impacts are ranked as 
L (cost increase $<\,$\SI{5}{\%}, schedule delay $<\,$\num{2} months), 
M (\SIrange{5}{25}{\%} and 2--6 months, respectively) and 
H ($>\,$\SI{20}{\%} and $>\,$2 months, respectively)}

\newduneabbrev{lbls}{LBLS}{laser beam location system}
{Auxiliary calibration system providing an independent location measurement of the ionization laser beams direction}

\newduneabbrev{lsst}{LSST}{Large Synoptic Survey Telescope}{8.4 m telescope with 3.2G-pixel camera that will start taking data in 2023}
\newduneabbrev{ska}{SKA}{Square Kilometer Array}{International radio telescope array planned to start data-taking in 2027}
\newduneabbrev{hyperk}{HyperK}{Hyper Kamiokande}{260 kt water Cerenkov neutrino detector to begin construction at Kamiokande in 2020}
\newduneword{lhcb}{LHCb}{LHC experiment dedicated to forward physics}
\newduneword{belleii}{Belle II}{B-factory experiment now running at KEK}

 \newduneabbrev{ldm}{LDM}{light-mass dark matter}{Refers to dark matter particles with mass values much lower than the electroweak scale, specifically below the 1~GeV level}
 
\newduneabbrev{bnv}{BNV}{baryon-number violating}{Describing an interaction where \gls{baryonnumber} is not conserved}

\newduneword{bugey}{Bugey}{Neutrino experiment that operated at the Bugey nuclear power plant in France}

\newduneword{minosplus}{MINOS$+$}{The successor to the \gls{minos} experiment, utilizing the same detectors and beam line, but operating at higher beam energy tune than \gls{minos}, parasitic with \gls{nova}}

\newduneword{baryonnumber}{baryon number}{A quantity expressing the total number of baryons in a system minus the number of antibaryons}

\newduneword{np04}{NP04}{CERN North Area hadron beamline used for the \gls{sp} test beam run}

\newduneword{ua1}{UA1}{UA1 (Underground Area 1) was a particle detector at \gls{cern}'s  Super Proton Synchrotron (SPS). It ran from 1981 until 1990, when the SPS was used as a proton-antiproton collider, searching for traces of W and Z particles in collisions. (CERN) The UA1 dipole magnet was reused in the NOMAD experiment and currently provides the magnetic field for the \gls{t2k} ND280 detector}

\newduneword{ssc}{SSC}{The Superconducting Super Collider was to be a huge underground ring complex beneath the area near Waxahachie, Texas, USA, that would have been the world’s most energetic particle accelerator. It was begun in 1990, but canceled by the U.S. Congress in 1993 (scientificamerican.com Oct 2013)}

\newduneword{daphne}{DAPHNE}{Detector electronics for Acquiring PHotons from NEutrinos is a custom-developed warm front-end waveform digitizing electronics module derived from the readout system developed at Fermilab for the Mu2e experiment}
 
\newduneword{nersc}{NERSC}{National Energy Research Computing Facility at \gls{lbnl}}

  \newduneword{integoff}{integration office}{The office that incorporates the onsite team responsible for coordinating integration and installation activities at SURF}

\newduneabbrev{sma}{SMA}{SubMiniature version A}{Connector interface for coaxial cables
with a screw-type coupling mechanism}

\newduneword{kloe}{KLOE}{KLOE is a $e^+ e^-$ collider detector spectrometer operated at DAFNE, the $\phi$-meson factory at Frascati, Rome. In DUNE it will consist of a \SI{26}{cm} Pb+scintillating fiber ECAL surrounding a cylindrical open detector region that is  \SI{4.00}{m} in diameter and \SI{4.30}{m} long. The ECAL and detector region are embedded in a \SI{0.6}{T} magnetic field created by a \SI{4.86}{m} diameter superconducting coil and a \SI{475}{tonne} iron yoke}

\newduneword{ro}{review office}{An office within the \gls{integoff} that organizes reviews }

\newduneabbrev{doecd}{CD}{critical decision}{The U.S. DOE's Order 413.3B outlines a series of staged project approvals, each of which is referred to as a critical decision (CD)}

\newduneabbrev{lbnfspac}{LBNF SPAC}{LBNF Strategic Project Advisory Committee}{A committee charged by the host laboratory director to provide expert, independent advice on significant issues and strategies related to LBNF project organization, management, and risks}

\newduneabbrev{sand}{SAND}{System for on-Axis Neutrino Detection}{The beam monitor component of the near detector that remains on-axis at all times and serves as a dedicated neutrino spectrum monitor}

\newduneword{4850l}{4850L}{The depth in feet (1480 m) of the top of the cryostats underground at SURF; used more generally to refer to the DUNE underground area. Called the ``4850 level'' or ``4850L''}

\newduneabbrev{cvmfs}{{\tt cvmfs}}{CernVM File System}{The CernVM File System provides a scalable, reliable and low-maintenance software distribution service. It was developed to assist High Energy Physics (HEP) collaborations to deploy software on the worldwide-distributed computing infrastructure used to run data processing applications}

\newduneabbrev{mrb}{{\tt mrb}}{Multi Repository Build System}{A Fermilab developed build system based on \dword{cmake} that allows development and builds of code from multiple repositories}

\newduneabbrev{cmake}{{\tt cmake}}{Cmake}{CMake is an open-source, cross-platform family of tools designed to build, test and package software}

\newduneabbrev{ups}{{\tt ups}}{Unix Product Support}{UPS is a Fermilab developed package manager that supports installation of multiple versions of a product and multiple builds per version.}
\makeglossaries

\begin{document}

\parskip 0 pt


\title{Computing for the DUNE Long-Baseline Neutrino Oscillation Experiment}

\author{\firstname{Heidi} \lastname{Schellman for the DUNE collaboration} \inst{1,2}\fnsep\thanks{\email{heidi.schellman@oregonstate.edu}}}
\institute{Department of Physics, Oregon State University, Corvallis OR, 97331 \and
Scientific Computing Division, Fermilab, Batavia, IL, 60555}

\abstract{
This is a talk given at Computers in High Energy Physics in Adelaide, South Australia, Australia in November 2019.  It is partially intended to explain the context of DUNE Computing for computing specialists.  

The \dword{dune}  collaboration consists of over 180 institutions from 33 countries. The experiment is in preparation now with commissioning of the first 10kT fiducial volume Liquid Argon TPC expected over the period 2025-2028 and a long data taking run with 4 modules expected from 2029  and beyond. 

An active prototyping program is already in place with a short test beam run with a 700T, 15,360 channel prototype of single-phase readout at the neutrino platform  at CERN in late 2018 and tests of a similar sized dual-phase detector scheduled for mid-2019.    The 2018 test beam run was a valuable live test of our computing model.  The detector produced raw data at rates of up to ~2GB/s.  These data were stored at full rate on tape at CERN and Fermilab and replicated at sites in the UK and Czech Republic.  In total 1.2 PB of raw data from beam and cosmic triggers were produced and reconstructed during the six week test beam run.


Baseline predictions for the full DUNE detector data, starting in the late 2020's are 30-60 PB of raw data per year.   In contrast to traditional HEP computational problems, DUNE's Liquid Argon TPC data consist of simple but very large (many GB) 2D data objects which share many characteristics with astrophysical images.  This presents opportunities to use advances in machine learning and pattern recognition as a frontier user of High Performance Computing facilities capable of massively parallel processing. 
}
\maketitle

\section{Introduction}

The \dword{dune}  will begin running in the late 2020's.  The goals of the experiment include 1) to studying neutrino oscillations using a beam of neutrinos from Fermilab in Illinois to the Homestake mine in Lead, South Dakota, 2) studying  astrophysical neutrino sources and rare processes and 3) understanding  the physics of neutrino interactions in matter.   In this talk I will concentrate on the neutrino oscillation and supernova capabilities of the experiment and the ways that they drive computing. 

The neutrino beam from Fermilab will consist almost entirely of muon-type neutrinos when produced.  Neutrinos are known to come in (at least) 3 flavors which can be distinguished by their interactions - electron type neutrinos produce electrons when they interact via charged currents, muon neutrinos, muons and tau neutrinos, tau particles.  But these flavors do not correspond to fixed mass states.  All 3 flavors of neutrinos are mixtures of mass states, much as  light in the $x$ direction  can be considered a superposition of  $x^\prime$ and $y^\prime$ polarizations along  alternate axes rotated by 45 degrees.  When neutrinos propagate through space, it is the mass state that sets their wavelength and if the neutrino goes far enough, the multiple mass states  corresponding to the initial flavor state will get out of phase.  When the mixture is later probed about its flavor, it  may give a different answer than the neutrino that started out. This phenomenon is neutrino oscillation and has been shown to exist in multiple experiments since it was first confirmed in 1998\cite{Kajita2006}.

\begin{figure}[h]
    \centering
\includegraphics[height=6cm]{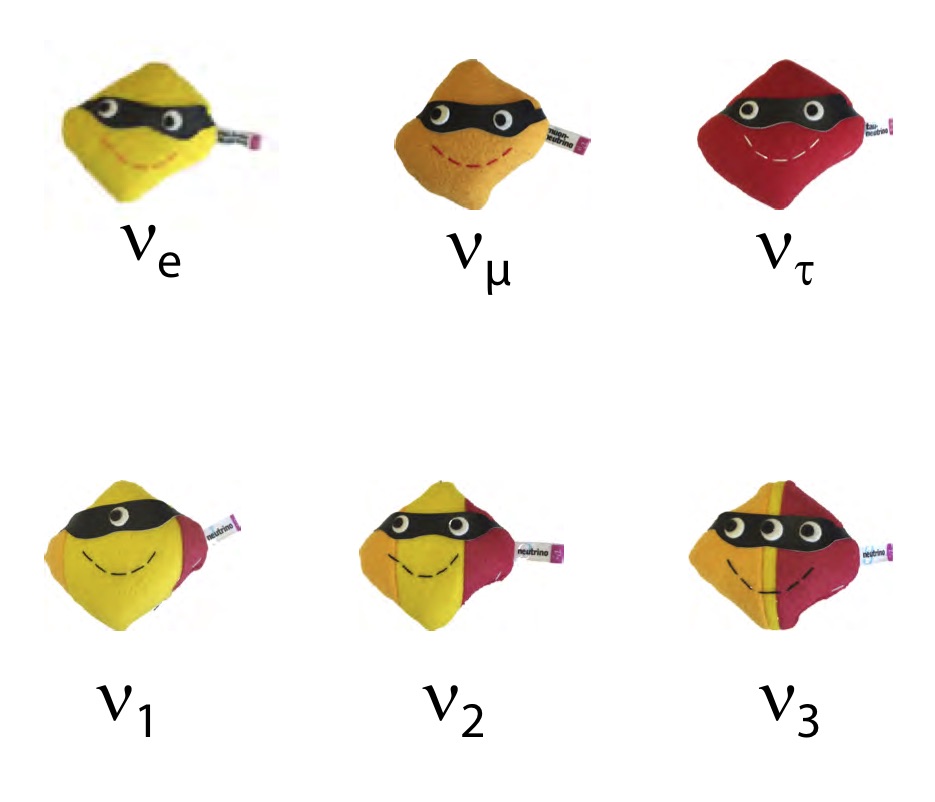}
    \caption{Illustration of the neutrino flavor and mass states.  The mass states are a superposition of the flavor states.  Courtesy the particlezoo.net.}
    \label{fig:neutrinos}
\end{figure}

\begin{figure}[h]
    \centering
\includegraphics[trim={0cm 0.6cm 2.5cm 0.7cm},clip,height=6cm]{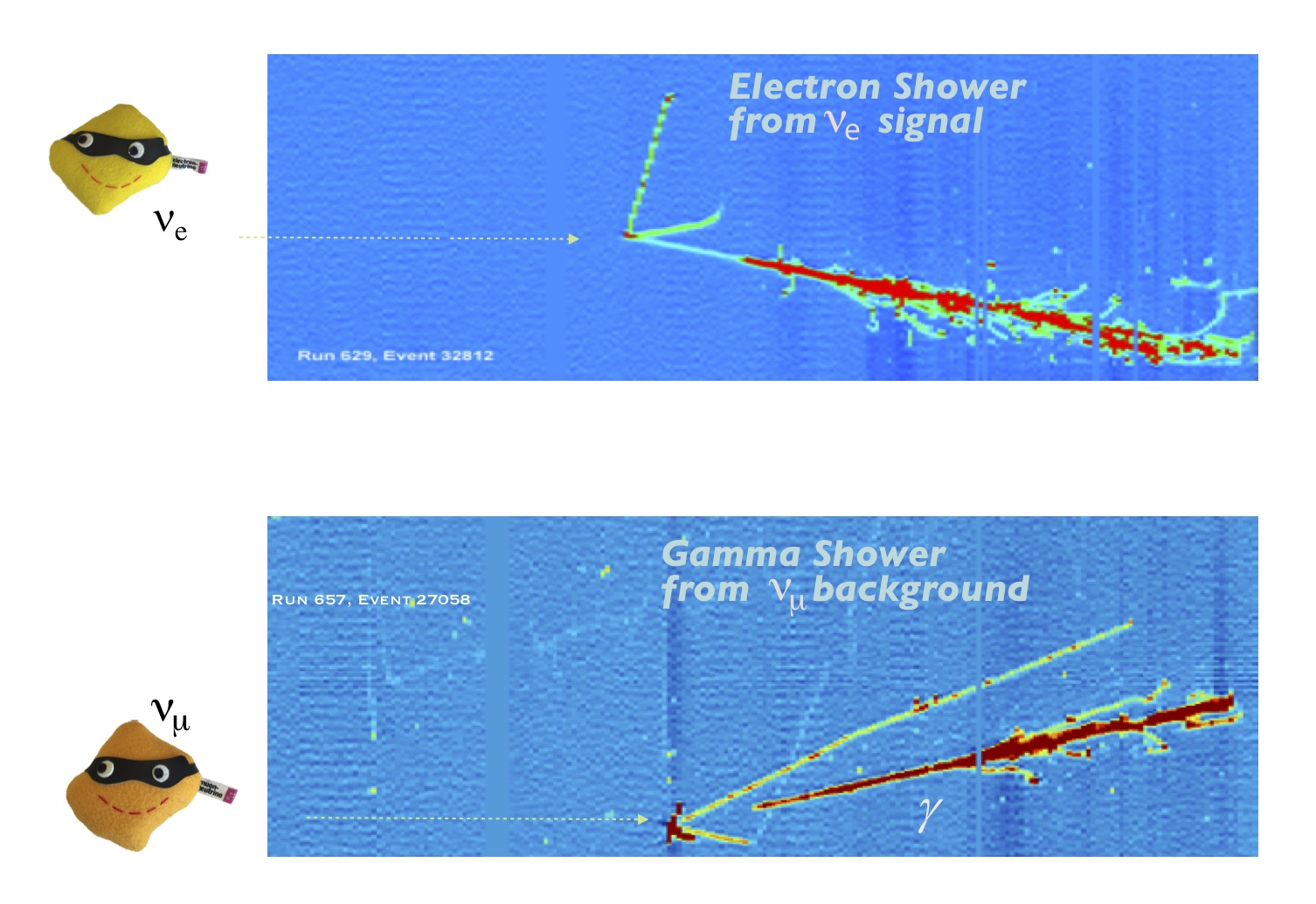}    \caption{Electron neutrino appearance signal (top) and background (bottom) as seen in the ArgoNeut experiment\protect{\cite{Acciarri:2016sli}}.  In the true appearance signal, an electron is seen emerging from the primary vertex and then showering.  In the background interaction, a muon neutrino enters and  produces a final state muon and photons, which propagate some distance before showering.}
    \label{fig:Argoneut}
\end{figure}

DUNE,  in particular,   wishes to understand the conversion of muon neutrinos created in Illinois into electron neutrinos at a \dword{fd} in the Homestake mine in South Dakota and compare that conversion rate between neutrino and anti-neutrino beams. The location of the \dword{fd} and energy of the neutrino beam were chosen to maximize the oscillation effect.   A difference in the conversion rate for neutrinos and anti-neutrinos could be evidence for matter-anti-matter asymmetry in the neutrino sector, a phenomenon called CP violation.  

To make these measurements, we need to be able to distinguish electron neutrino interactions appearing in the muon neutrino beam from the dominant muon neutrino interactions one would expect in the absence of oscillations.  Doing this requires a very large detector, as neutrino interactions are intrinsically rare, but an extremely  fine grained one as well.  Noble liquid time projection detectors, which read out large transparent volumes of liquid by drifting electrons from interactions to charge detectors through strong electric fields, have the needed capabilities of extremely large scale and fine-grained resolution. The proposed DUNE far-site detector will instrument four  $14\times12 \times58$ meter volumes of Liquid Argon with readout granularity of 0.5 cm.  The detectors will be located 4850 ft below the surface to lower the rate of cosmic rays traversing the detector by orders of magnitude and thus allow sensitivity to very low energy solar and astrophysical neutrinos as well as the higher energy neutrinos produced at Fermilab.

The neutrino beam from Fermilab will be pulsed approximately once/second 24 hrs/day during running periods with of order 15 million pulses per year.  Because neutrinos interact  extremely rarely, we expect to detect of order 7,500  neutrino interactions/year in each of 4 10~kT detector modules located at the \dword{fd} site in South Dakota.

Construction of the detector halls and infrastructure for the large 10 kT fiducial volume \dword{fd} modules is starting now, as are design and construction of detector readout modules.  A full \dword{tdr} for the program has recently been completed and is available in references \cite{Abi:2020wmh, Abi:2020evt, Abi:2020oxb, Abi:2020loh}.
The  \dword{dune} neutrino oscillation experiment will receive beam late in this decade with commissioning of the data acquisition systems for the first far detector module expected to start in 2025-26.  

\begin{figure}\label{DUNESchematic}
\includegraphics[height=0.35\textwidth]{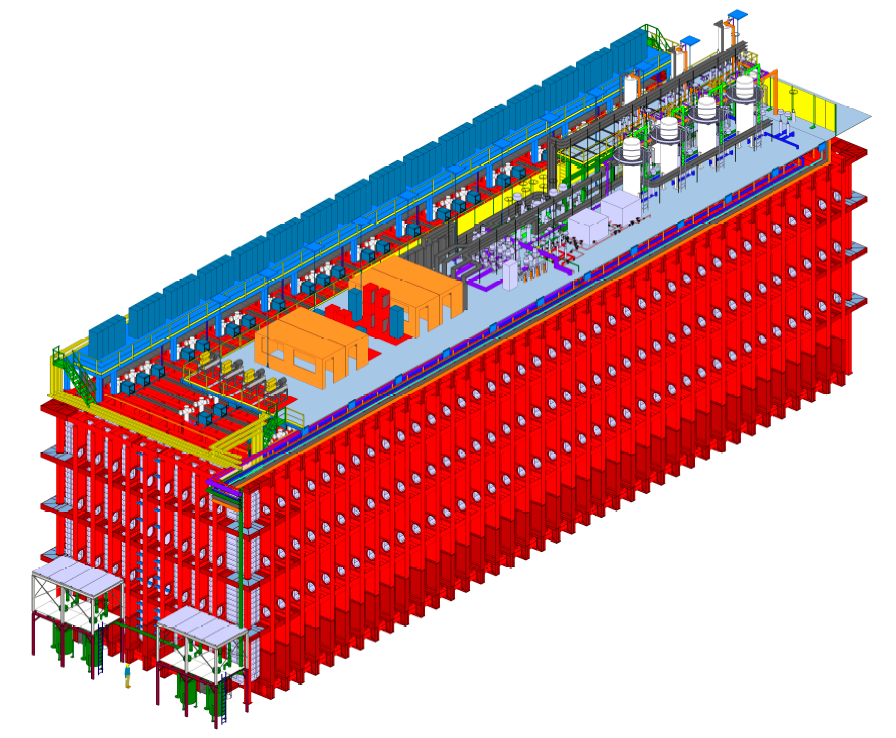}
\includegraphics[height=0.35\textwidth]{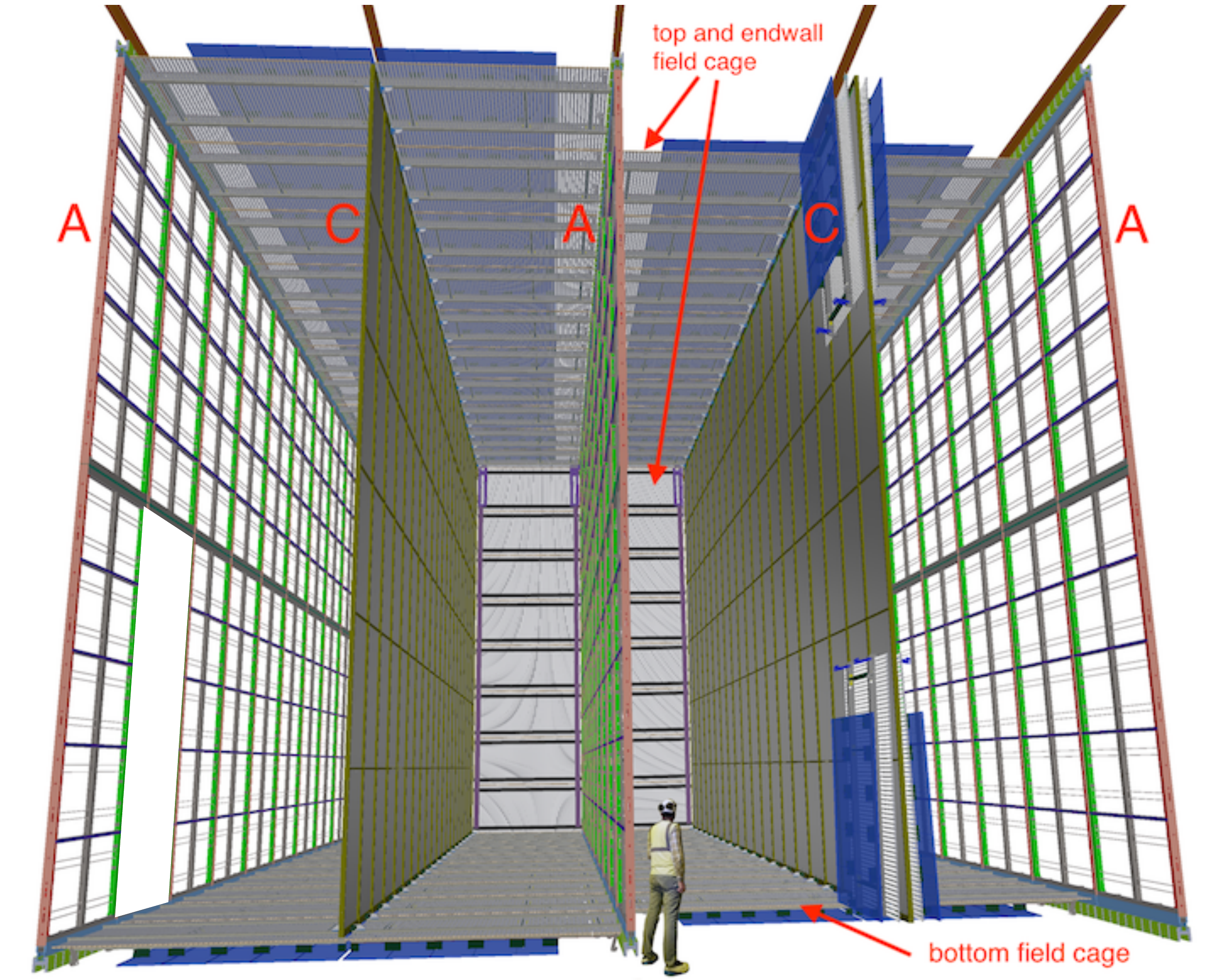}
\caption{Left) A far detector cryostat that houses a 10 kT \dword{fd} module. The figure of a person indicates the scale.  Right) A 10kt  \dword{dune} \dword{fd} \dword{spmod}, showing the alternating 58 m long (into the page), 12 m high anode (A) and cathode (C) planes, as well as the field cage that surrounds the drift regions between the anode and cathode planes. The blank area on the left side was added to show the profile of a single \dword{apa}.}

\end{figure}

\section{ProtoDUNE tests at CERN}

Building an experiment of this size requires an extensive period of prototyping.   The Argoneut\cite{Acciarri:2018myr}, MicroBooNE\cite{microboone} and ICARUS\cite{icarus} collaborations have demonstrated the capabilities of large liquid Argon \dword{tpc}s for neutrino detection on scales between 1 and 500 ton fiducial mass.  In preparation for the \dword{dune} experiment, a campaign testing proposed DUNE components in 700 ton detectors in the EHN1 hadronic test beam was launched at CERN in 2018.  Both single-phase and dual-phase prototypes were constructed and tested. 

\subsection{\dword{pdsp}}
\begin{figure}[h]
    \centering
\includegraphics[trim={0cm 0.6cm 2.5cm 0.7cm},clip,height=8cm]{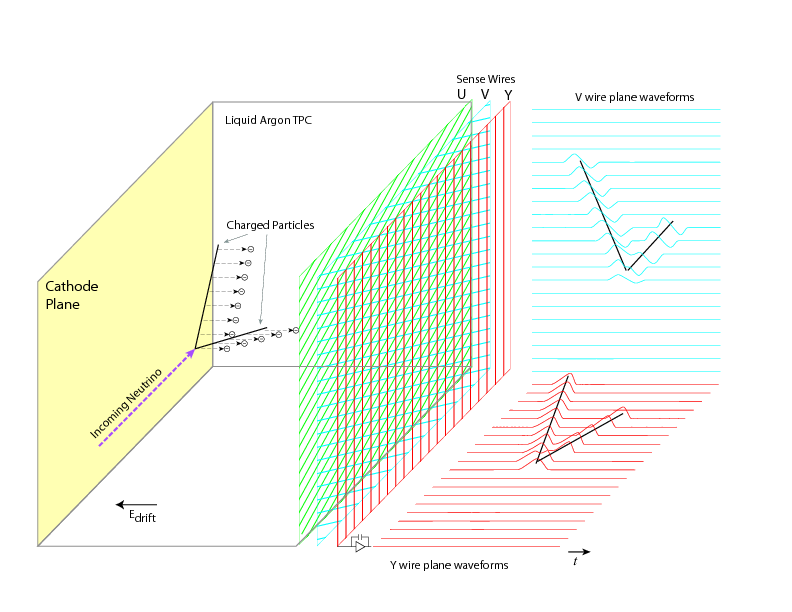}
    \caption{Diagram  from  \protect{\cite{ Acciarri:2017sde}}  illustrating the signal formation in a LArTPC with three wire planes~\cite{Acciarri:2016smi}. For simplicity, the signal in the first U induction plane is omitted in the illustration. }
    \label{tpcconcept}
    \end{figure}

The \dword{pdsp} experiment began taking data at CERN in late 2018.  \dword{pdsp} uses single-phase technology where ionization electrons are collected directly from the liquid argon. The readout system consists of  Anode Plane Assemblies (\dword{apa})s which each have 3 layers of wires arranged in different directions. Each layer contains 800-1200  wires spaced 0.5 cm apart. Electrons drift from the original interaction in the Argon, through a strong electric field, to the wire planes and induce signals.  The location in the plane of hit wires gives one coordinate, the time the signal takes to drift to the wire from the original interaction measures a second coordinate.  The third coordinate is derived by combining information from overlaps of signals in the 3 different wire layers.  Signals are amplified electronically and then digitized.  Figure \ref{tpcconcept} illustrates the operation of a generic \dword{lartpc}.

The \dword{pdsp} detector consists of a 700 ton volume of liquid argon with a cathode plane in the center and 3 \dword{apa}s mounted on  each  edge of the liquid volume.  The drift distance is  3 m with  a nominal voltage of 180kV  across that distance.  Each \dword{apa} has 2560 channels and each channel reads out a 12-bit \dword{adc} every 0.5 $\mu$sec.   For \dword{pdsp} the readout time appropriate for a 3 m drift was set to 3 msec, resulting in 6000 12-bit samples per channel.  The total data size for six \dword{apa}s is thus 140 MB with additional header and data from photon and external tagging systems bringing the nominal event size up to around 180 MB.  Lossless compression of the \dword{tpc} readout was implemented in the data acquisition system, resulting in a final compressed event size of around 75 MB. 

The test beam ran at rates of up to 25 Hz over a period of 6 weeks at beam momenta between 0.5 and 7 GeV/c.  Time of flight and Cherenkov counters provided beam flavor tagging.  Around 8M total `physics' events were written, with around 3M having beam tag information.  In total  850 TB of raw test beam data were written, along with one PB of commissioning and cosmic data. These data were successfully cataloged and written to storage at both CERN and Fermilab at rates of up to 2 GB/sec.   

\begin{figure}
\includegraphics[width=0.49\textwidth]{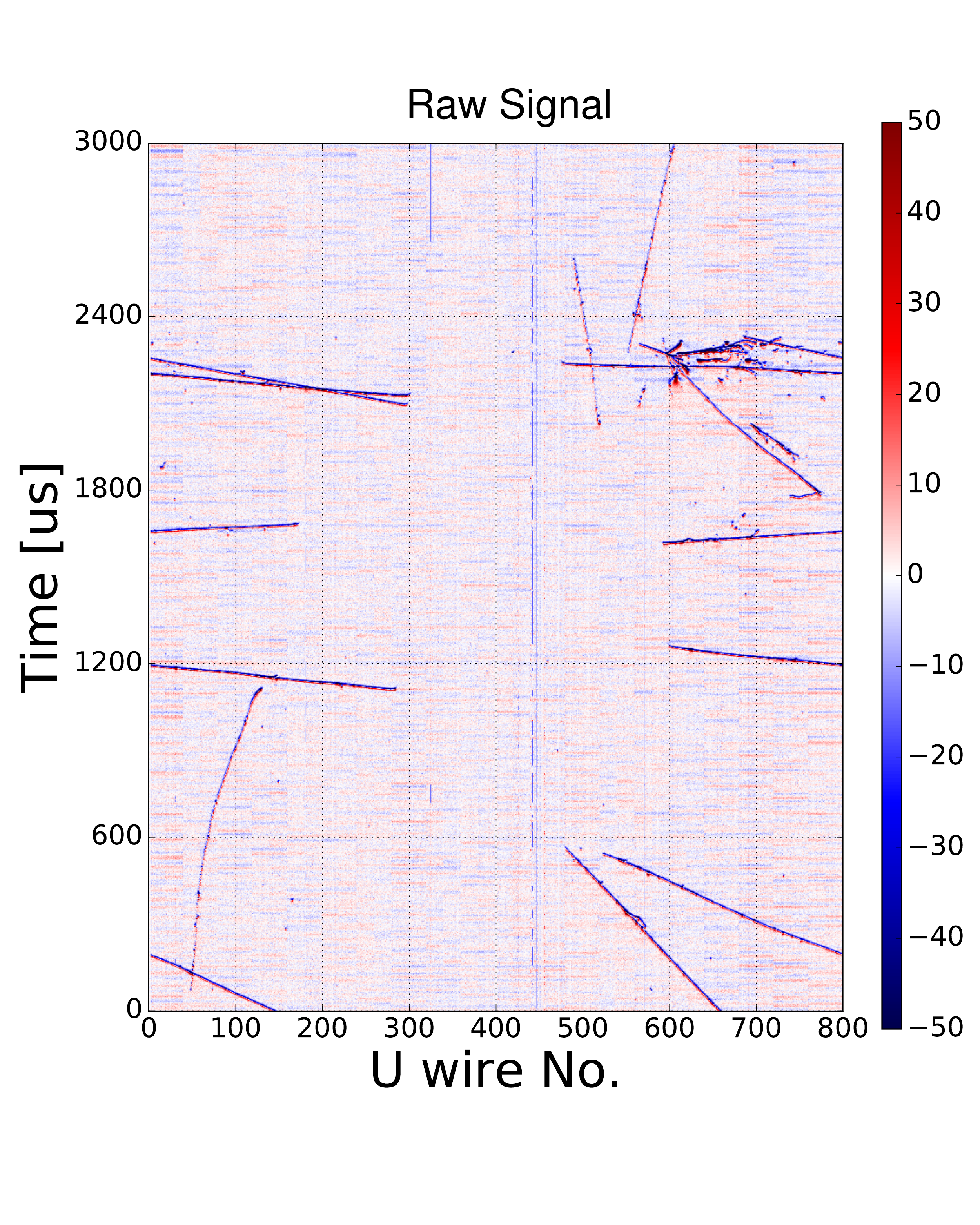}
\includegraphics[width=0.49\textwidth]{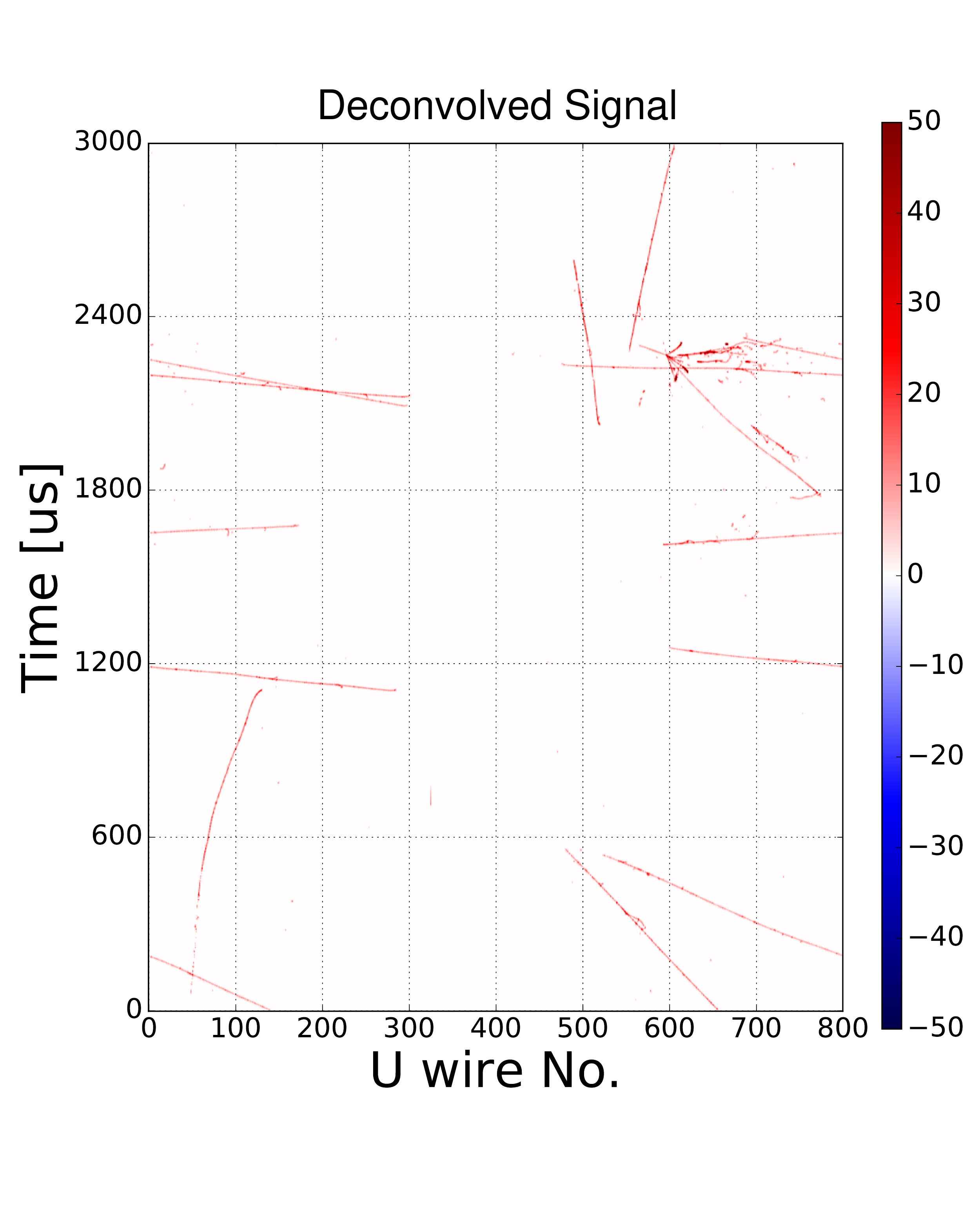}
\caption{Comparison of raw (left) and deconvolved induction U-plane signals (right) before and after 
the signal processing procedure from a \dword{pdsp} event. The bipolar shape with red (blue) color representing
positive (negative) signals is converted to the unipolar shape after the \twod deconvolution.}
\label{deconvolution}
\end{figure}

\begin{figure}

\includegraphics[width=0.9\textwidth]{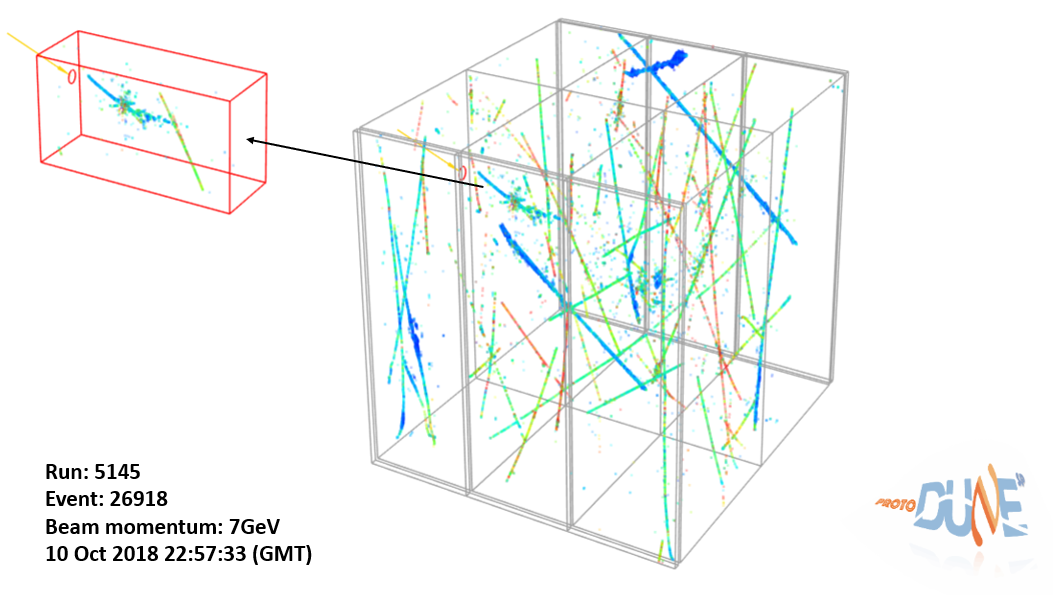}
\caption {The \dword{pdsp} detector (gray box) showing 
the direction of the particle beam (yellow line on the very far left) and the outlines of the six \dword{apa}s. Cosmic rays
can be seen throughout the white box, while the red box highlights the beam region of interest with an interaction of the 7 GeV beam. 
The \threed points are obtained using the Space~Point~Solver reconstruction algorithm. }
\label{wire-cell-bee}
\end{figure}

\begin{figure}
\includegraphics[width=0.8\textwidth]{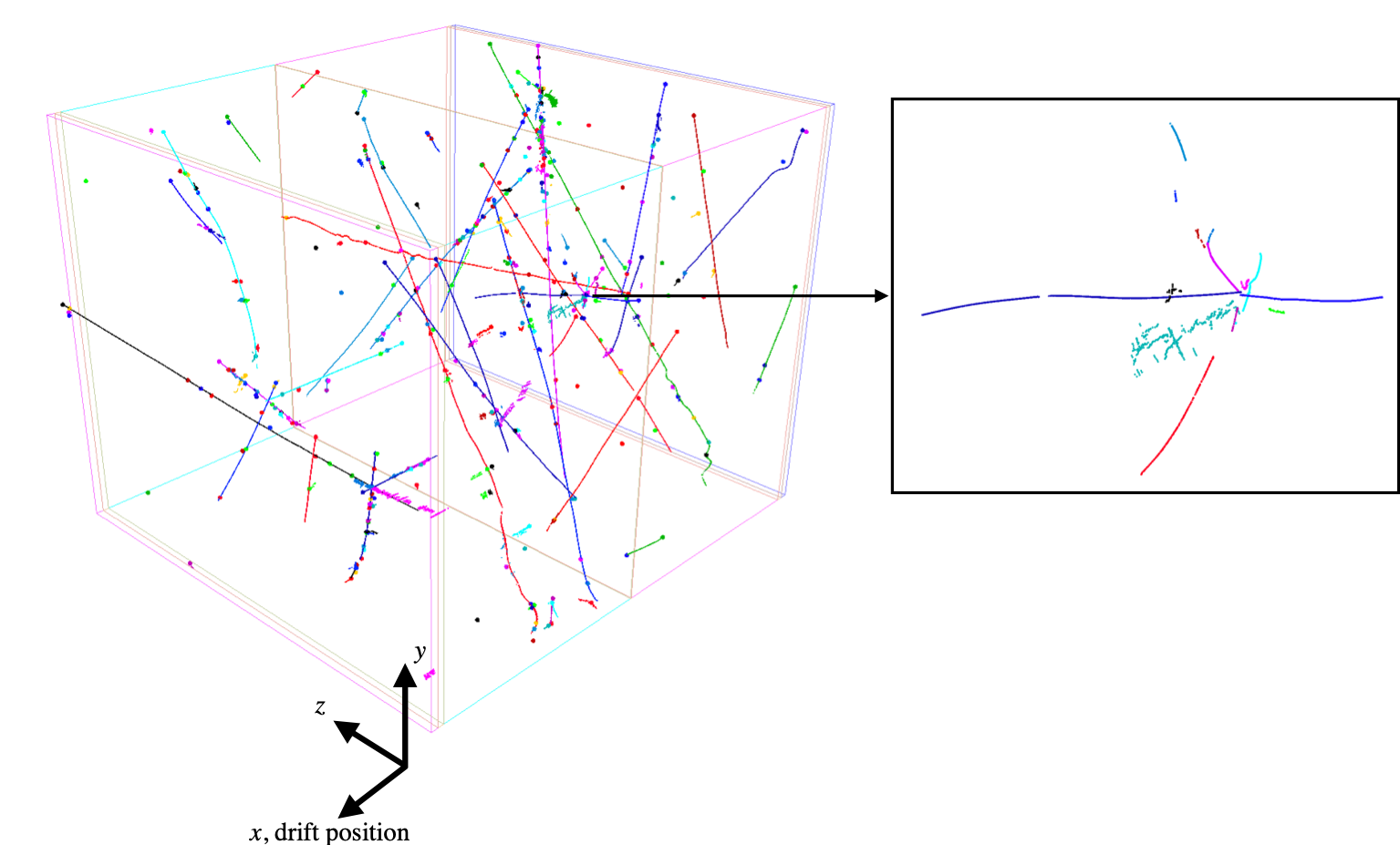}
\caption{Pandora \protect{\cite{Acciarri:2017hat}} reconstruction of cosmic rays and beam interaction in a \dword{pdsp} event. The left side of the figure shows the full detector volume with all interactions, including cosmic rays and the right side shows the identified beam interaction.}
\label{pandora}
\end{figure}

Thanks to significant prior effort in the \dword{lar} computing and algorithms community, reconstruction software was ready to go and the first reconstruction pass began soon after data taking started and was complete within two weeks of the end of data taking.  Those results were extremely useful in demonstrating the capabilities of the detector and summarized in Volume II of the \dword{tdr}\cite{Abi:2020evt}.  A second pass, with improved treatment of instrumental effects ranging from stuck bits to 2-D deconvolution to correction for space charge effects was completed in late 2019. 
Figure \ref{deconvolution} illustrates the signal processing stage of reconstruction, where raw ADC signals have noise and stuck bits removed and are then deconvoluted to yield gaussian hit candidates. Figures \ref{wire-cell-bee} and \ref{pandora} illustrate full pattern recognition and event reconstruction. 

While \dword{lartpc}s benefit from fine granularity and a uniform detector medium, diffusion, argon purity, fluid flow and the build up of space charge in the active medium can all introduce distortions into the detector response.  These effects have all been simulated and tested in the \dword{pdsp} data. 

\begin{figure}
\begin{center}
\includegraphics[height=0.5\textwidth]{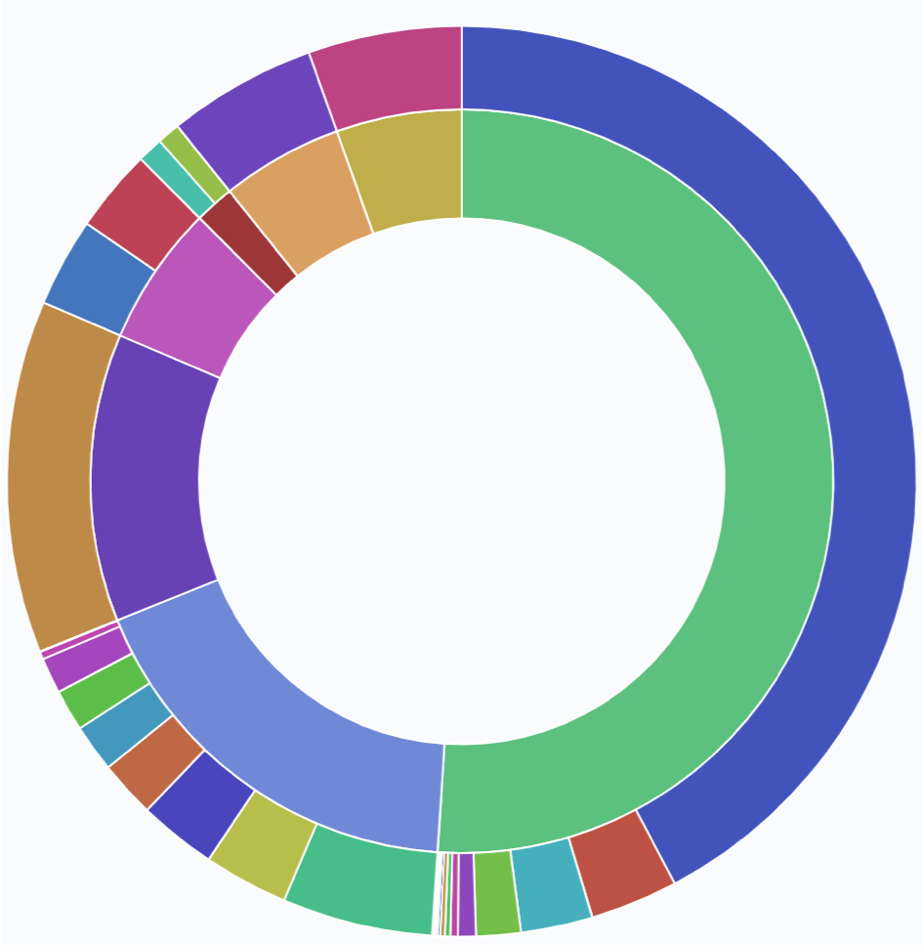}
\includegraphics[height=0.5\textwidth]{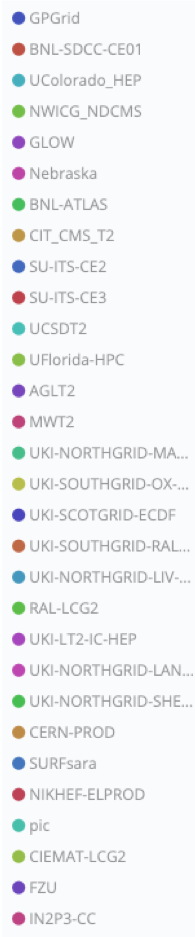}
\caption{Reconstruction processing distribution amongst sites for DUNE production in calendar 2019.  The inner circle shows national contributions while the outer circle shows individual site contributions.}
\label{sites}
\end{center}
\end{figure}

Compressed raw input event records were of order 75 MB in size and took 500-600 seconds to reconstruct, of which around 180 sec was signal processing and the remainder high level reconstruction dominated by 40-60 cosmic rays per readout.  Memory footprints ranged between 2.5 and 4 GB.  Output event  record sizes were reduced to 22 MB by dropping the raw waveforms after hit finding.   Reconstruction campaigns took of order 4-6 weeks (similar to the original data taking) and utilized up to 15,000 cores on \dword{osg}/\dword{wlcg} resources.  Job submission was done through the POMS\cite{poms} job management system developed at Fermilab. POMS supports submissions to FNAL dedicated resources and selected OSG and WLCG sites.  Figure \ref{sites} shows the distribution of wall hours used for reconstruction in 2019. 

For reconstruction, data were streamed via {\tt xrootd}\cite{Behrmann:2011zz} from {\tt dcache} storage at Fermilab to the remote sites. Despite individual processing jobs taking 15-30 hrs to complete, network interruptions rarely caused job failures.

\subsection{\dword{pddp}}

\begin{figure}
\label{dpevent}
\includegraphics[width=0.85\textwidth]{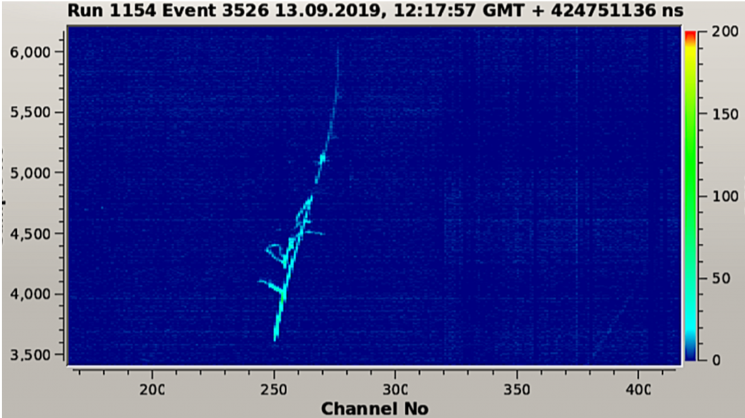}
\caption {Cosmic ray data from the dual phase prototype}
\end{figure}

The \dword{pddp} detector began taking data using cosmic rays in August 2019.  Thanks to preceding data challenges, those data have been successfully integrated into the full data cataloging and reconstruction chains and are now being reconstructed as they arrive.   The \dword{pddp} technology locates the readout systems above a thin layer of argon gas above the liquid argon surface.  This gas layer allows an external electric field to accelerate the electrons and produce gas amplification.  The result is a substantial increase in signal-to-noise in the resulting signals, at the cost of longer electron drifts from the bottom of the liquid volume.  Figure \ref{dpevent} illustrates early data from \dword{pddp}. 

\subsection{Conclusions from prototype tests}
\dword{protodune} prototype runs are ongoing and will continue through beam tests in 2021-22 at \dword{cern}.  Data cataloging, movement and storage techniques were tested before the start of the \dword{pdsp} and \dword{pddp}  runs and were able to handle the full rate of the experiments.   Reconstruction algorithms were also in place on time and were able to produce early results that led to increased understanding of the detector and improved calibrations for a second iteration.  These tests also identified some deficiencies in our infrastructure, including incomplete schemes for the transmission of configuration and conditions information between the hardware operations and the offline computing.  The test beam runs have been extremely valuable in allowing us to determine which variables are important to transmit and in designing improved systems for gathering and storing that information. 

An additional run of both \dword{pdsp} and \dword{pddp} is planned for 2021 and 2022, allowing further developing and testing of our computing infrastructure before the full detector comes online in the late 2020's.

\section{On to full DUNE}

The full DUNE \dword{fd} will begin with one single phase module to be installed at Homestake starting in the middle of  this decade.  High intensity neutrino and anti-neutrino beams should arrive after a year or so of commissioning of the detector and \dword{lbnf} beamline.  This first module will largely resemble a scaled up version of \dword{pdsp} with 150 \dword{apa}s distributed 2-deep at the center and long edges of the cryostat.   The argon volume will be $15\times14\times62$ m$^3$ with a fiducial mass of 10kT.  Table \ref{volumes} summarizes the expected event rates and data volumes for one such detector module.  Additional detector modules, likely one dual-phase, another single-phase and  one with novel technology will be added.  For now, we assume that data volumes and rates coming from other technologies will be less than or equal to the single-phase values.

The detectors should be sensitive to neutrino interactions and radioactive decays above an energy threshold of order 5 MeV.  Unambiguous triggering may require a somewhat higher threshold  to avoid false triggers due to $^{39}$Ar decays but beam interactions in the 500-10,000 MeV range should have almost perfect detection efficiency. Sophisticated triggering algorithms should also allow standalone detection of astrophysical sources, including higher energy solar neutrinos and supernova candidates. 

The data rates will be dominated by 4,500 cosmic rays expected per module/day.  These events are vital for monitoring and aligning the detector.  The next most significant source of events will be calibration campaigns with radioactive and neutron sources and lasers.  In all cases, the goal is to gather data from the full volume of the detector with as fine a granularity as possible. 

Beam interactions themselves are expected to be quite rare, occurring in only 1/2000 beam gates.  Extraction of oscillation parameters will require both the powerful electron background rejection discussed in the previous section and precise calibration of the energy scale of the experiment, hence the much larger calibration samples.

 \begin{table}[htp]

\begin{center}
\begin{tabular}{|l |r r r |}
\hline
Process & Rate/module & \qquad size/instance &\qquad  size/module/year\\
\hline
Beam event & 41/day & 6 GB&47 TB/year\\
Cosmic rays &4,500/day&  6 GB& 9.7 PB/year\\
Supernova trigger& 1/month& 115 TB& 1.4 PB/year\\
Calibrations&2/year&750 TB& 1.5 PB/year\\
\hline 
Total& & &12.9 PB/year\\
\hline
\end{tabular}
\end{center}
\label{volumes}
\caption{Data sizes and rates for different processes in each far detector module.  Uncompressed data sizes are given. As readouts will be self-triggering an extended 5.4 ms readout window is used instead of the 3ms for the triggered \dword{pdsp} runs.  We assume beam uptime of 50\% and 100\% uptime for non-beam science.These numbers are derived from references \protect{\cite{bib:docdb16028} }and \protect{\cite{bib:docdb14983}}.}
\end{table}%

Overall, bottoms-up estimates yield data volumes of around 13 PB/year/module.  Lossless compression should reduce this volume and additional modules will likely increase these rates.  A maximum rate of 30PB/year across all modules and modes of operation has been specified.  We will note that 30 PB/year is  an average of 1.3 GB/sec, less than the rates already demonstrated for protoDUNE acquisition and storage.  In principle, at 2.5 CPU sec/MB of compressed input, 2000-3000 cores could keep up with these data rates  but this throughput must be maintained over many years.   In addition, supernova candidates may require bursts of  much higher acquisition and processing rates. Table \ref{tab:exec-comp-bigpicture-es} summarizes the computational characteristics expected for \dword{fd} data.

\subsection{Supernova candidates}

\begin{figure}
\begin{center}
\includegraphics[height=0.3\textwidth]{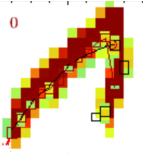} \hskip 1 in
\includegraphics[height=0.5\textwidth]{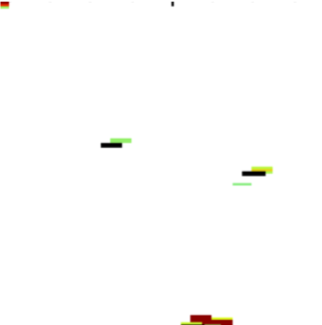}
\caption{Left) a charged current interaction of a 30 MeV energy electron neutrino in the DUNE Far Detector.  Right) a neutral current excitation and de-excitation of an Ar nucleus by a  10 MeV neutrino.}
\label{blips}
\end{center}
\end{figure}

Supernova candidates pose a unique problem for data acquisition and reconstruction.  Supernova physics in DUNE is discussed in some detail in the \dword{tdr}\cite{ Abi:2020evt} and only summarized here. A classic core-collapse supernova 10 kpc away would be expected to yield around 3,000  charged-current electron neutrino interactions across 4 detector modules.  The oscillation physics is not fully understand and can result in significant modulations of the event rates for different neutrino types  over the few tens of seconds of the burst.  DUNE's fine-grained tracking should allow significant pointing power with the most optimistic scenario of four modules and high electron neutrino fraction yielding pointing resolutions of less than 5 degrees.   Figure \ref{blips} illustrates simulated signatures of supernova neutrino interactions in the far detector. The ability to produce a reasonably fast pointing signal would be extremely valuable to optical astronomers doing followup, especially if a supernova was in a region where dust masks the primary optical signal.   The need to be alert to supernovae and to quickly transfer and process the data imposes significant requirements on triggering, data transfer and reconstruction beyond those imposed by the more regular beam-based oscillation physics.   For example, a compressed supernova readout of all four modules will be of order 184 TB in size and take a minimum of 4 hrs to transfer over a 100 Gbs network,  and then take of order 130,000 CPU-hrs for signal processing at present speeds.  If processing takes the same time as transfer, a peak of 30,000 cores would be needed.

\begin{table}
\begin{center}
\begin{tabular}{|l  l |c       | l |}
\hline
Quantity&&\qquad Value \qquad&Explanation\qquad \qquad \\
\hline
{\bf Far Detector Beam:}&&&\\ 
&Single APA readout &41.5 MB& Uncompressed 5.4 ms\\ 
&Single APA readout &16.6 MB& $\times 2.5$ compression\\
&APAs per module& 150&\\
&Full module readout &6.22  GB& Uncompressed 5.4 ms\\ 
&Beam rep. rate&0.83 Hz&Untriggered\\  
Signal processing &CPU time/APA&40 sec&from MC/ProtoDUNE\\  
Signal processing &CPU time/input MB& 2.5 sec/MB& compressed input\\
&Memory footprint/APA&0.5-1 GB&ProtoDUNE experience\\  
\hline
{\bf Supernova:}&&&\\
&Single channel readout &300 MB& Uncompressed 100 s\\  
&Four module readout& 460 TB& Uncompressed 100 s\\  
&Trigger rate&1  per month&(assumption)\\
\hline
\end{tabular}
\caption{Useful quantities for computing estimates for \dword{sp}
readout. For  sparse \dword{fd} events, the pattern recognition phase, which scales with occupancy is expected to be substantially faster than the signal processing phase which scales with detector size.  }%
\label{tab:exec-comp-bigpicture-es}
\end{center}
\end{table}



\section{Comments and Conclusions}
This discussion has centered on the acquisition and fast processing of raw data from novel and extremely large liquid argon time projection chambers. Many other computing challenges lie ahead but were beyond the scope of this talk.  These include

\begin{description}
\item[{\bf Simulation:}] particle propagation in liquid argon is reasonably fast to simulate as there are not complicated volume boundaries to cross but simulating electron drift trajectories (and scintillation light trajectories) in a diffusive, electron absorbing, moving medium immersed in a non-uniform  electric field remains a challenging computational challenge. 
\item[{\bf Near detectors:}] a suite of near detectors are needed to characterize the neutrino beam as it originates at Fermilab.  These detectors are still being developed but will introduce a large number of differing detector technologies.  While individual interactions are likely to be much smaller than readouts of the far detectors, the beam cycle is of order 1 Hz and each readout will contain multiple cosmic ray and beam interactions.
\item[{\bf Data analysis:} ] The small (order 100 ) group of \dword{pdsp} and \dword{pddp} and \dword{dune} developers and analyzers have successfully analyzed the beam and cosmic ray data and performed simulations needed to produce the physics sections of the \dword{tdr}.  We expect analysis of the full experiment to involve many more individuals and much more data.  A campaign of training for new users and design of a suite of efficient analysis tools is needed.  We have initial prototypes based on NOvA and MicroBooNE analysis. 
\end{description}

Fortunately, \dword{dune} is able to take advantage of the huge and heroic developments in software and computing made for the Intensity Frontier and LHC experiments over the past decade.  We have demonstrated that, even with preliminary versions of our tools and algorithms, we can quickly reconstruct and analyze data from large liquid argon TPC's at full rate. We look forward to an exciting and fruitful next decade. 

\section*{Acknowledgements}
Schellman's research is supported by the  US National Science Foundation under grant 1806849 and through a joint appointment with the Fermilab Scientific Computing Division.
Fermilab is a  U.S. Department of Energy, Office of Science, HEP User Facility managed by Fermi Research Alliance, LLC (FRA), acting under Contract No. DE-AC02-07CH11359.

\bibliography{tdr-citedb}


\end{document}